\documentclass[11pt]{article}
\usepackage{preprint}

\usepackage{amsmath, bm}
\usepackage{mathtools}   % for \coloneqq
\usepackage{dsfont} % for indicator function in FSS
\usepackage{amssymb} % in preprint.sty
\usepackage[capitalise, noabbrev]{cleveref}
\usepackage{graphicx}
\graphicspath{{figures/}{../figures/}}
\usepackage{wrapfig}
\usepackage{tikz} % for annotating raster figures, see the sample forecasts in the SI
\usepackage{float} % provides the [H] "put it exactly here" figure placement
\usepackage{longtable} % for SI tables that run past a single page

\usepackage{soul}
\usepackage{todonotes}

\definecolor{accent}{HTML}{004074}   % AGU navy

\definecolor{notsig}{gray}{0.62}
\newcommand{\nsig}[1]{\textcolor{notsig}{#1}}

\usepackage[authoryear, round]{natbib}
\newcommand{\meanci}[1]{Lines show the mean over #1~forecasts and shading the
95\% confidence interval.}
\newcommand{\medianci}[1]{Lines show the median over #1~forecasts and shading
the 95\% confidence interval.}

\newcommand{\state}{\bm{x}}
\newcommand{\pred}{\hat{\bm{x}}}
\newcommand{\obs}{\bm{y}}
\newcommand{\opred}{\hat{\bm{y}}}
\newcommand{\model}{\mathcal{M}}
\newcommand{\params}{\bm{\theta}}

\usepackage{lineno}
\usepackage{titletoc} % partial (SI-only) table of contents, see below

\newcommand{\papershorttitle}{Nested-EAGLE}
\newcommand{\papertitle}{%
    Bridging short- and medium-range\\
    weather forecasting with machine learning%
}

\newcommand{\papertitleplain}{%
    Bridging short- and medium-range weather forecasting with machine learning%
}
\newcommand{\paperauthorsplain}{%
    Timothy A. Smith, Mariah Pope, Sergey Frolov, Brett Basarab,
    Daniel Abdi, Paul Madden, Isidora Jankov%
}
\newcommand{\paperkeywords}{%
    machine learning weather prediction, limited area model, nested model,
    HRRR, GFS, medium-range forecasting%
}

\newif\ifpapermetaset
\newcommand{\setpapermeta}[1]{%
    \ifpapermetaset\else
        \papermetasettrue
        \hypersetup{%
            pdftitle    = {#1},
            pdfauthor   = {\paperauthorsplain},
            pdfsubject  = {Machine learning weather prediction},
            pdfkeywords = {\paperkeywords},
        }%
    \fi
}

\newcommand{\setmaintitle}{%
    \shorttitle{\papershorttitle}%
    \title{\papertitle}%
    \setpapermeta{\papertitleplain}%
}

\newcommand{\setsititle}{%
    \shorttitle{\papershorttitle\ Supporting Information}%
    \title{Supporting Information for\\``\papertitle''}%
    \setpapermeta{%
        Supporting Information for
        \textquotedblleft\papertitleplain\textquotedblright%
    }%
}

\authors{%
    Timothy A. Smith\textsuperscript{1,2,3,\dag,*},
    Mariah Pope\textsuperscript{4},
    Sergey Frolov\textsuperscript{1,*},
    Brett Basarab\textsuperscript{1, 5},
    Daniel Abdi\textsuperscript{5, 6},
    Paul Madden\textsuperscript{5, 6},
    Isidora Jankov\textsuperscript{6}
}

\affiliations{%
    \textsuperscript{1}Physical Sciences Laboratory (PSL), National Oceanic and
        Atmospheric Administration (NOAA), Boulder, CO, USA\\
    \textsuperscript{2}Nansen Environmental and Remote Sensing Center,
        Bergen, Norway\\
    \textsuperscript{3}Bjerknes Center for Climate Research,
        Bergen, Norway\\
    \textsuperscript{4}Earth Prediction Innovation Center (EPIC)\\
    \textsuperscript{5}Cooperative Institute for Research in Environmental Sciences
        (CIRES) at the University of Colorado Boulder, Boulder, CO, USA\\
    \textsuperscript{6}Global Systems Laboratory (GSL), National Oceanic and
        Atmospheric Administration (NOAA), Boulder, CO, USA\\
    \textsuperscript{\dag}Now at Nansen Environmental and Remote Sensing Center,
        and Bjerknes Center for Climate Research, Bergen, Norway
}

\email{timothy.smith@nersc.no, sergey.frolov@noaa.gov}

\setmaintitle

\begin{document}
\maketitle

\begin{abstract}
    The National Oceanic and Atmospheric Administration (NOAA) employs
    independent prediction systems for distinct forecast products.
    While some
    separation is practical, we argue that combining short- and medium-range
    weather into a single prediction system would provide the public with a
    useful distillation of global weather and its impacts.
    To this end, we present Nested-EAGLE
    (Experimental Artificial intelligence Global and Limited-area Ensemble):
    a 0.25$^\circ$ global weather model with a 6~km refinement
    over the Contiguous United States (CONUS).
    The model achieves significantly lower mean-squared error in near-surface and low-level
    quantities over CONUS compared to NOAA's Global Forecast System and
    High-Resolution Rapid Refresh (HRRR),
    while remaining competitive throughout the rest of the global atmosphere.
    We show that the skill gains for near-surface fields stem
    from incorporating high-resolution regional analysis data into training through the nesting process.
    Forecasts of precipitation amounts are less skillful than those from HRRR,
    owing to deterministic training.
    However, we show that Nested-EAGLE provides the most accurate forecasts of
    storm locations at longer leads, despite blurred extrema.
    Our results motivate future work to extend the skill gains beyond CONUS and
    improve precipitation representation.
\end{abstract}
% ----------------------------------------------------------------

\section{Introduction}

% NOAA prediction suite
The National Oceanic and Atmospheric Administration (NOAA) regularly broadcasts
weather and climate forecasts to the public for a variety of applications,
including, for example,
severe thunderstorms,
hurricanes,
10~day weather,
and seasonal outlooks of temperature and precipitation.
Most of these forecast products are produced by distinct systems tuned to the
application at hand, each with its own model.
That is, there is not a singular forecast system at NOAA,
even though the underlying modeling framework is unified for many cases
\citep{jacobs_open_2021}.
Of course, some separation is both practical and necessary.
For example, it does not make sense to operate a moving nest hurricane model outside of
hurricane season.
However, we postulate that
forecasters would benefit from having a single prediction system developed for multiple
applications, as this would reduce the array of results to synthesize when
providing a forecast.

% Hone in on SRW and MRW
We are specifically focused on jointly predicting short-range weather,
spanning hours to a few days,
and medium-range weather, which extends to about two weeks.
At NOAA, short-range weather is captured by the High-Resolution Rapid Refresh
(HRRR): a 3~km resolution Limited Area Model (LAM) covering the Contiguous United States
(CONUS).
HRRR forecasts are currently initialized every hour by the
analysis (i.e., initial conditions) coming from the
13~km resolution
Rapid Refresh (RAP) data assimilation system, and extend either 18 or 48~hours into the
future \citep{dowell_high-resolution_2022}.
Deterministic medium-range weather is handled by NOAA's Global Forecast
System (GFS):
a global model initialized every 6~hours by the analysis from NOAA's Global Data Assimilation
System (GDAS), with forecasts released at 0.25$^\circ$ resolution \citep{noaa_emc_nfs}.
GFS and HRRR are not entirely independent from one another, since HRRR and RAP use GFS
and GDAS data as boundary conditions.
However, forecasters must use the products separately for the information that
they provide.
For example, GFS data are often used
to provide a two-week outlook
and to inform the synoptic-scale weather context,
while HRRR data are used to isolate local
impacts for only up to a one- or two-day lead.

% ML offers the opportunity to do this
Developments in Machine Learning Weather Prediction (MLWP)
offer new opportunities to join short- and medium-range weather forecast
systems.
In recent years, several global MLWP models
%trained on the
%ERA5 \citep{hersbach_era5_2020}
have been developed
\citep{bi_accurate_2023,lam_learning_2023,chen_fuxi_2023,chen_operational_2025,bodnar_foundation_2025,lang_aifs_2024,sadeghi_tabas_gfs-powered_2025}
that are competitive with or outperform the gold
standard in traditional medium-range weather forecasting:
the IFS
(Integrated Forecast System)
from
ECMWF
(European Centre for Medium-Range Weather Forecasts).
Since the development of global, medium-range MLWP models, several regional weather
emulators have been produced as well
\citep[e.g.,][]{oskarsson_graphbased_2023,adamov_building_2025,flora_wofscast_2025,pathak_kilometer-scale_2026,abdi_hrrrcast_2026}.
By construction, regional MLWP systems inherit the high spatial resolution of
the traditional LAM analyses and forecasts they are trained on.
However, while ML models can benefit from new perspectives (for instance
\citealt{adamov_building_2025} discuss how ML LAMs can blend
forecasting and downscaling tasks),
many existing regional MLWP models still provide distinctly separate
information from their global counterparts.

\begin{figure}[tbp]
    \centering
    \includegraphics[width=\textwidth]{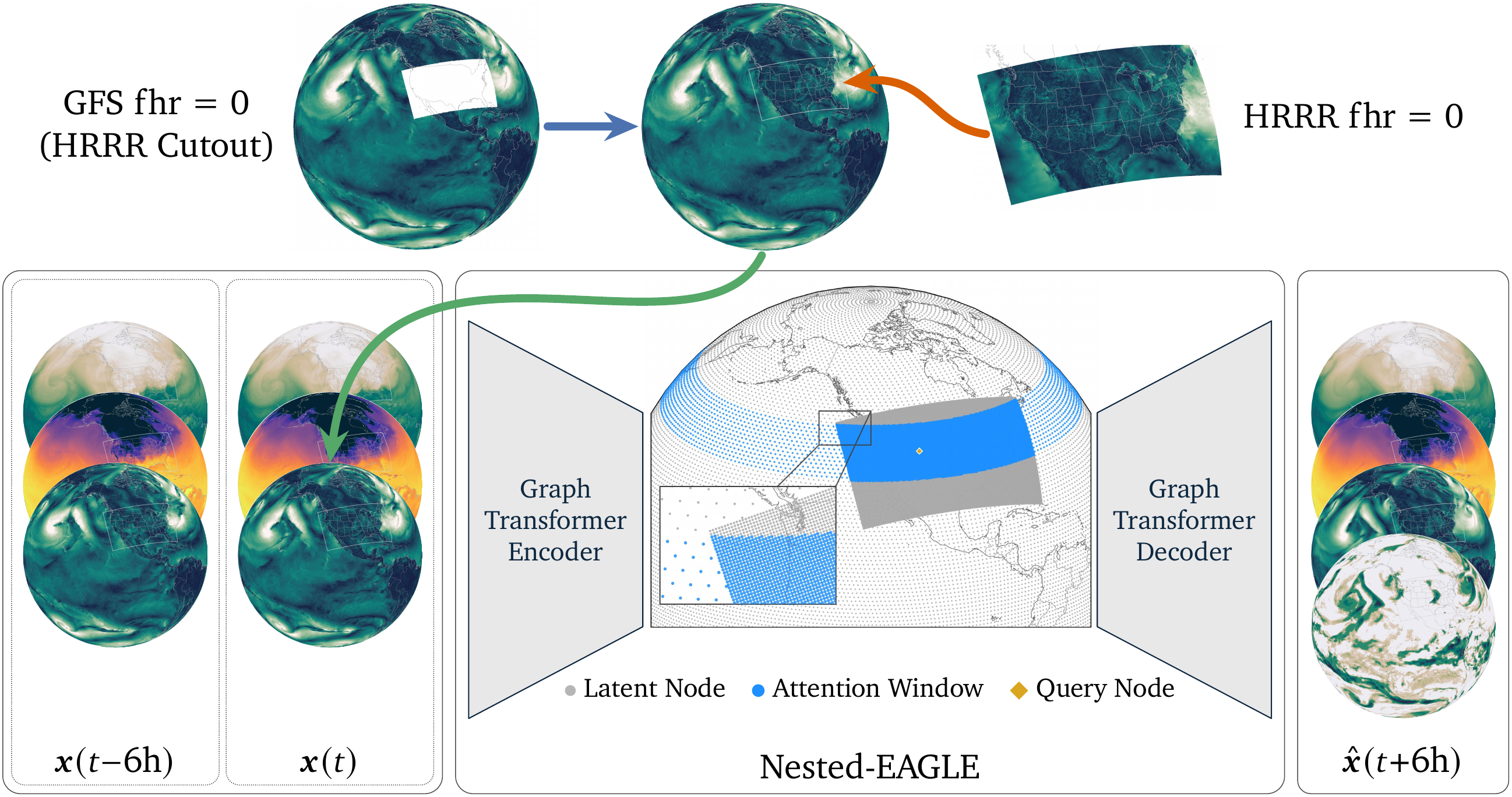}
    \caption{
        \textbf{Illustration of Nested-EAGLE.}
        The model operates on a state space composed of HRRR data over
        CONUS, conservatively regridded to 6~km resolution, and GFS data
        everywhere else on the globe at 0.25$^\circ$ resolution.
        For a given timestamp $t$, Nested-EAGLE is initialized with this nested
        state at $t$ and $t-6$~hours, and makes a prediction 6~hours into the
        future.
        The model is parameterized by a deep neural network that consists of a
        graph-transformer encoder and decoder, and a sliding-window transformer
        processor.
        The processor acts on a latent mesh that is essentially a coarsened
        version of the nested data space, with relatively higher resolution over
        CONUS than elsewhere.
        The figure shows the latent mesh nodes in gray, and an example
        attention window passing over the high-resolution CONUS region given a
        query node over Boulder, Colorado.
        Additionally, the figure illustrates that Nested-EAGLE diagnoses some
        quantities, like accumulated precipitation shown here, which are not
        provided as inputs.
        For this figure, we used $t$ = 0600 UTC 8 March 2023.
        See \cref{subsec:nested-eagle-design} for more design details and
        \cref{si:sample-forecasts} for more snapshots of several variables from a sample
        forecast.
    }
    \label{fig:nested_illustration}
\end{figure}

% Nesting / stretched grid
In contrast to the typically separated global and regional model development
pipelines, \citet{nipen_regional_2026} developed Bris: an ``all-in-one'' MLWP model
that has high (2.5~km) resolution over Scandinavia, but still captures the rest
of the globe at $\sim$31~km.
After a pretraining phase on ERA5 data \citep{hersbach_era5_2020},
the model is trained on a combination of data from the LAM and global forecast
system archives, resulting in state-of-the-art prediction skill out to 2.5~days
at a fraction of the computational cost (ignoring training costs)
compared to Met Norway's traditional
forecast system, the Meteorological cooperation Ensemble Prediction System
(MEPS).
Crucially, Bris shows that a combined MLWP system can provide high-quality
forecasts for a region of interest.

% Now introduce Nested-EAGLE
Here, we build on the work by \citet{nipen_regional_2026} to develop
Nested-EAGLE
(Experimental Artificial intelligence Global and Limited-area Ensemble):
an MLWP model with a 6~km mesh over CONUS, nested
inside a $0.25^{\circ}$ grid.
See \cref{fig:nested_illustration}
for an illustration and \cref{si:sample-forecasts} to visualize more fields from a
sample forecast.
Similar to how Bris was designed using MEPS and
IFS data, we trained Nested-EAGLE on NOAA's HRRR and GFS archives,
using HRRR data over CONUS and GFS elsewhere.
Nested-EAGLE is currently a deterministic model with a 6~hour time step.
As such, we designed the model to highlight how nesting operational
regional analysis and forecast data into a single global MLWP model
improves skill.
We show that Nested-EAGLE can bridge the short and medium range
by extending the evaluation protocol considered by
\citet{nipen_regional_2026}, evaluating out to 15~days both in and out of
the target region.
We then compare perturbation experiments to show that combining the datasets via
nesting is only beneficial if it is done during training, providing some
intuition for how the model compares to existing single-resolution emulators.

Nested-EAGLE does not yet resolve the convective scales that are important for
short-range and extreme weather prediction.
We show the model's current limitations in representing
precipitation, due to the known blurring effect of deterministic training at
6~hour time steps using a Mean-Squared Error (MSE) loss function
\citep[e.g., as noted by][]{lam_learning_2023}.
We dissect the precipitation evaluation to show that Nested-EAGLE
struggles to represent extreme precipitation amounts, but is generally
successful in predicting storm locations, thereby isolating areas of improvement
for future work.

\section{Prognostic Forecast Skill}
\label{sec:prognostic_skill}

We evaluate the performance of Nested-EAGLE
by comparing its forecast skill against the relevant physics-based modeling
frameworks from NOAA: GFS and HRRR (see \cref{subsec:baseline-datasets} for
dataset details).
We also compare Nested-EAGLE against a 0.25$^\circ$ resolution global MLWP
model that we call ML-GFS-Base.
We developed ML-GFS-Base almost identically to Nested-EAGLE, but only trained it on
GFS data in order to isolate the impact of incorporating HRRR data via nesting.
See \cref{subsec:nested-eagle-design} and \cref{subsec:baseline-design} for
design details on Nested-EAGLE and ML-GFS-Base, respectively.

In our prognostic skill evaluation,
e.g., for 2m temperature,
we compare 15~day forecasts initialized every 30~hours during the test period,
February 2024--January 2025, evenly sampling the
diurnal and seasonal cycles throughout the year, resulting in 293~samples per
lead time.
We evaluate forecasts from each model in terms of Root-Mean-Squared Error
(RMSE) against in situ (conventional) observations, after bilinearly
interpolating them to the observation locations.
See \cref{subsec:prognostic-evaluation} for more details on the evaluation
protocol.

\subsection{Forecast Skill Over CONUS}
\label{subsec:conus_skill}

\begin{figure}[t]
    \centering
    \includegraphics[width=\textwidth]{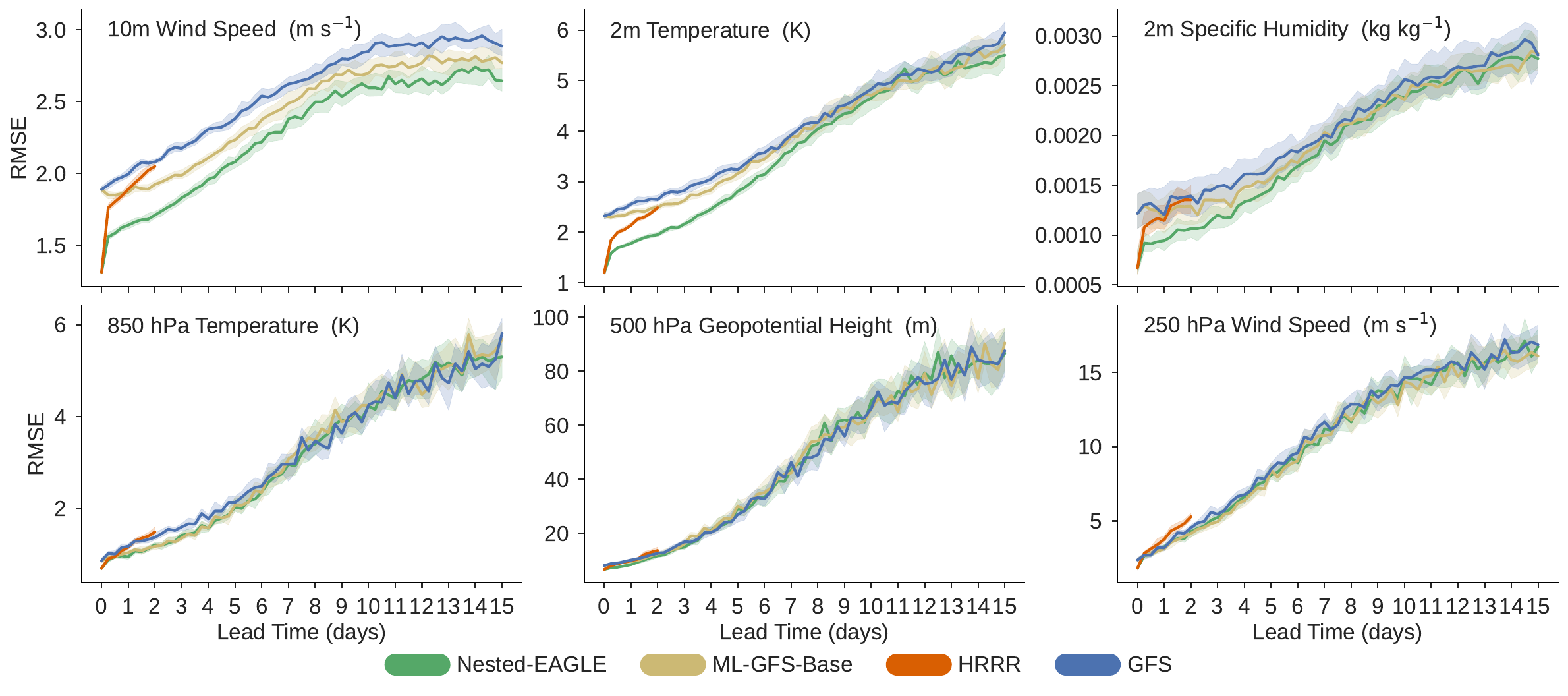}
    \caption{
        \textbf{
            RMSE against in situ observations over CONUS during the test period.
        }
        The model forecasts were first conservatively regridded to the same 6~km resolution
        Lambert conformal conic
        projection that Nested-EAGLE employs over CONUS, then bilinearly interpolated to the
        observation locations.
        See \cref{sifig:conus-rmse} for an evaluation of more quantities and
        vertical levels, and
        \cref{si:lead-time-equivalence} for a
        quantification of skill differences in terms of lead time.
        \medianci{293}
    }
    \label{fig:lam_rmse}
\end{figure}

\cref{fig:lam_rmse} shows that Nested-EAGLE generally has the lowest RMSE over
CONUS, especially for near-surface quantities, and remains competitive
throughout the atmosphere to 15~days.
More specifically, \cref{fig:lam_rmse} shows RMSE for Nested-EAGLE, ML-GFS-Base, and
GFS for 15~days of lead time, and HRRR for 2~days of lead time.
In order to add some perspective to the RMSE plots, we describe the skill based on the
lead time gap, $\tau_\text{gap}$, between Nested-EAGLE and the other baselines,
which defines the point at which median RMSE is approximately equal, i.e.,
\begin{equation}
    \text{Nested-EAGLE Median RMSE}(\Delta t+\tau_{\text{gap}}) \simeq
    \text{Baseline Median RMSE}(\Delta t) \, .
    \label{eq:skill-gap}
\end{equation}
Here $\tau_\text{gap}>0$ indicates the additional lead time over which
Nested-EAGLE has lower error.
For more details and tabulated numbers, see \cref{si:lead-time-equivalence}.
At one day of lead time, the skill gap between Nested-EAGLE and GFS is about
78~hours for 10m wind speed and 2m temperature, and 60~hours for 2m specific
humidity.
For 2m temperature and specific humidity, the skill gap between GFS and Nested-EAGLE
closes as lead time increases, remaining significant until about 7.5 and 2~days,
respectively.
On the other hand, Nested-EAGLE maintains at least a 48~hour skill gap over GFS
throughout the 15~day evaluation period for 10m wind speed.

Comparing Nested-EAGLE and ML-GFS-Base for near-surface variables (top row of
\cref{fig:lam_rmse}) shows the main
benefit of incorporating HRRR data into training.
At one day of lead time, Nested-EAGLE leads ML-GFS-Base by 54 and 66~hours for
10m wind speed and 2m temperature, respectively, and maintains a statistically
significant lead until about 6.5~days of lead time.
For 2m specific humidity, the skill gap is large ($\tau_\text{gap}>3$~days) but short-lived,
as it is only significant to about 12~hours of lead time.

The bottom row of \cref{fig:lam_rmse} shows RMSE for a selection of fields
indicating skill across the atmospheric column; see \cref{sifig:conus-rmse} for
more quantities and vertical levels.
For these atmospheric variables, the ML models achieve nearly equal
skill.
At 850~hPa, Nested-EAGLE and ML-GFS-Base have a statistically significant
$\tau_\text{gap}\simeq 24$~hour skill gap over GFS for the first 1--2~days of lead
time.
Beyond 4~days of lead time, all reporting models produce skill that is statistically
indistinguishable.
Higher up in the atmosphere, for example as shown by 500~hPa geopotential
height and 250~hPa wind speed errors, Nested-EAGLE, ML-GFS-Base, and GFS show
comparable skill.

Nested-EAGLE shows the most significant improvements over all other models near
the surface purely because of differences between the
HRRR and GFS analysis states, which we represent with data at forecast hour
zero in our evaluation.
For near-surface variables, HRRR analysis leads the GFS analysis by a
significant margin, amounting to improvements of roughly 30--48\%
(30\% for 10m wind speed, 48\% for 2m temperature, and 45\% for 2m specific
humidity).
We presume that these near-surface improvements are due to several factors in
the HRRR data assimilation system, including:
\begin{itemize}
    \item higher spatial resolution,
    \item more frequent assimilation cycles (hourly versus 6~hourly),
    \item a more advanced land model, which can make better use of surface
observations, and
    \item better observational coverage, using more U.S.-centric station data.
\end{itemize}
On the other hand, analysis differences in the atmospheric fields shown in the bottom row
of \cref{fig:lam_rmse} are only around 20\%.
In other words, HRRR and GFS have similar atmospheric states, but HRRR
translates this representation into a significantly better near-surface state.
Importantly, the relative improvements near the surface are not due to scaling
the loss function to prioritize the surface and deprioritize the atmosphere,
for example, as is done by
GraphCast \citep{lam_learning_2023} and the Artificial Intelligence Forecasting
System \citep[AIFS;][]{lang_aifs_2024}.
In fact, we found this scaling to be either inconsequential or even detrimental
to skill across all variables;
see \cref{si:pressure-scaling} for more details.
Thus, Nested-EAGLE's skill gain for near-surface quantities is due to the
propagation of skill gains in the HRRR analysis, indicating the importance of
incorporating the data into training.

\subsection{Forecast Skill Over Europe}
\label{subsec:europe_skill}

\begin{figure}[t]
    \centering
    \includegraphics[width=\textwidth]{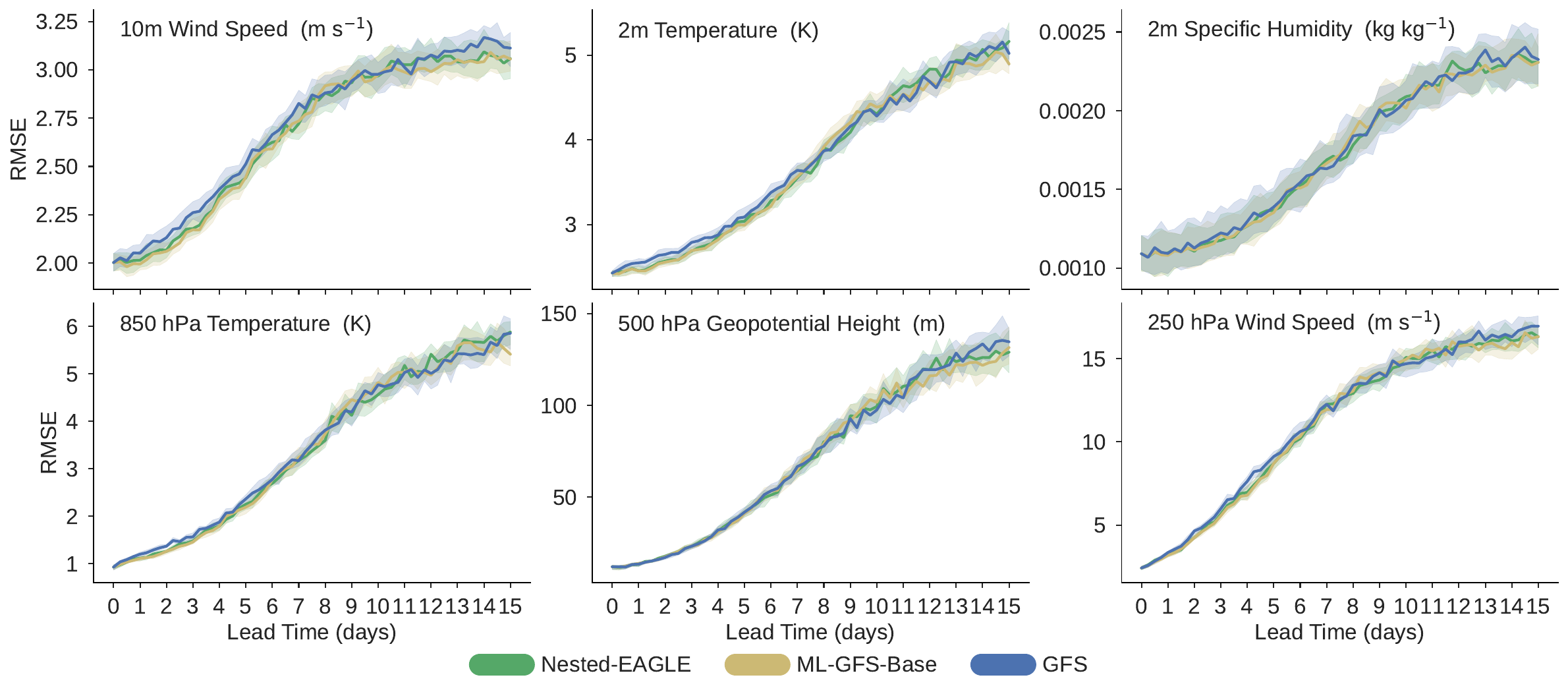}
    \caption{
        \textbf{
            RMSE against in situ observations over Europe during the test period.
        }
        The European region is defined as a simple latitude-longitude box from
        35$^\circ$N--75$^\circ$N and 25$^\circ$W--65$^\circ$E.
        All datasets operate on a 0.25$^\circ$ latitude-longitude grid in this
        region, so they are not regridded prior to observation-location interpolation.
        See \cref{sifig:europe-rmse} for an evaluation of more quantities and
        vertical levels.
        \medianci{293}
    }
    \label{fig:europe_rmse}
\end{figure}

\cref{fig:europe_rmse} indicates that outside of CONUS, our target region,
Nested-EAGLE does not carry over skill learned from the HRRR dataset, but
remains a reliable medium-range weather model.
The figure shows RMSE against in situ observations over Europe, comparing
Nested-EAGLE, ML-GFS-Base, and GFS.
Here, the two MLWP models show modest gains over GFS in 10m wind speed, 2m
temperature, and 850 hPa temperature.
Otherwise, all three models are statistically indistinguishable from each
other.
We suggest that the relatively small improvement over GFS in this region is due
to the limited 8~year training datasets used in our work.

We also interpret the statistical equivalence of Nested-EAGLE and ML-GFS-Base to
be a negative control.
While it would be ideal for Nested-EAGLE to carry over skill gains from the HRRR
dataset to the rest of the globe, our current framework provides no mechanism
for this to happen.
More specifically, the model is trained with a
supervised learning objective that encourages it to ``look like'' GFS data
outside of CONUS.
\cref{fig:europe_rmse} indicates that the model is not overtrained to its
target region; the skill gains over CONUS do not degrade skill elsewhere.

\subsection{Disentangling the Importance of Training Data and Inference-Time
Initial Conditions}
\label{subsec:ic_experiment}

The previous sections show that incorporating HRRR analysis into global MLWP model
training improves its representation of near-surface quantities over CONUS.
However, a fundamental limitation of Nested-EAGLE, and most of the existing MLWP models
from the broader community, is its reliance on an analysis state
that comes from traditional data assimilation systems.
% Leave this until discussion time
%generally provided
%publicly by operational centers like NOAA or ECMWF.
In our framework, which nests the HRRR analysis inside the GFS analysis, it is
not exactly clear \textit{how} the skill gain manifests.
Are the skill gains in Nested-EAGLE, relative to ML-GFS-Base, purely due
to the fact that we feed it
an improved initial state at inference time?
Or, are the improvements baked into
the model weights during training?

\begin{figure}[t]
    \centering
    \includegraphics[width=\textwidth]{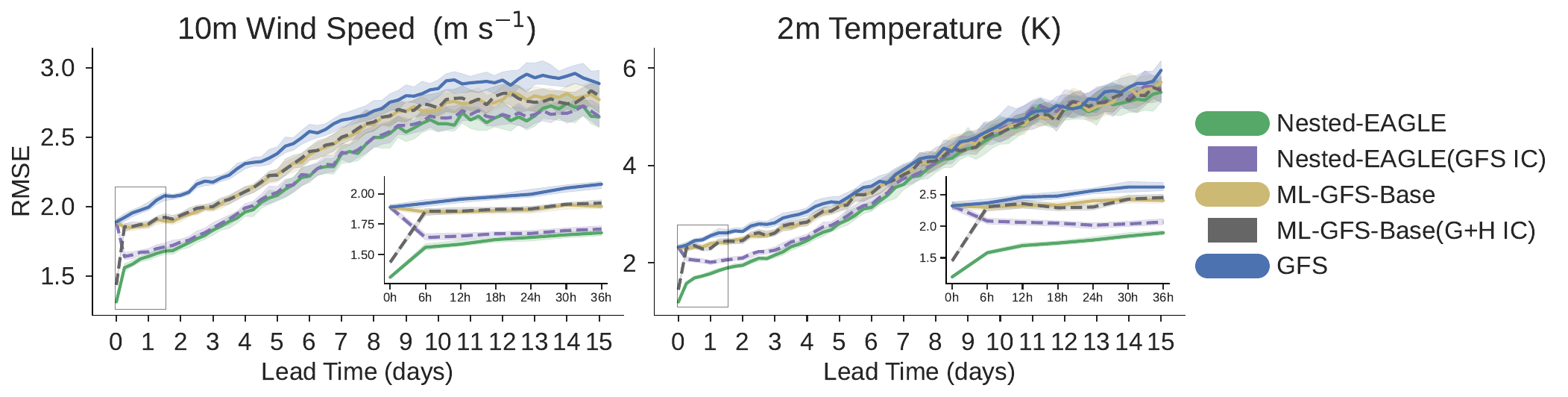}
    \caption{
        \textbf{
            RMSE over CONUS. Similar to \cref{fig:lam_rmse}, but with additional
            experiments using swapped initial conditions.
        }
        The additional purple dashed line shows the RMSE of Nested-EAGLE when it
        is initialized with GFS data only at inference time, which we call
        Nested-EAGLE(GFS~IC), where IC denotes the initial conditions.
        The gray dashed line shows the opposite: during inference ML-GFS-Base
        is fed initial conditions that have HRRR data over CONUS for
        ML-GFS-Base(G+H~IC), where G+H denotes combined GFS and HRRR initial
        conditions.
        For this figure, all datasets were conservatively regridded to a 0.25$^\circ$ resolution
        and evaluated over an approximate CONUS latitude-longitude box from
        20$^\circ$N--55$^\circ$N and 135$^\circ$W--50$^\circ$W.
        \medianci{293}
    }
    \label{fig:ic_experiment}
\end{figure}

\cref{fig:ic_experiment} shows that the improvements are learned during
training and carry over even with less accurate initial conditions.
Similar to the top row of \cref{fig:lam_rmse}, \cref{fig:ic_experiment} shows
near-surface forecast RMSE over CONUS from Nested-EAGLE, ML-GFS-Base, and GFS.
Additionally, \cref{fig:ic_experiment} shows the skill when Nested-EAGLE is
initialized with only GFS analysis at inference time (Nested-EAGLE(GFS~IC), purple dashed line).
In order to make the nesting mechanics work,
the GFS state was conservatively interpolated to the 6~km Lambert conformal conic
projection over the HRRR subdomain, and left ``as-is'' over the rest of the globe.
Conversely, \cref{fig:ic_experiment} also shows the skill of ML-GFS-Base,
when it is given HRRR initial conditions at
inference time (ML-GFS-Base(G+H~IC), gray dashed line).
Similarly, for the mechanics to work with the single-resolution model,
the HRRR state was conservatively coarsened to a 0.25$^\circ$
latitude-longitude projection, and swapped into the global grid as appropriate.

In \cref{fig:ic_experiment}, models initialized with swapped
analysis states start with the expected errors.
For example,
Nested-EAGLE(GFS~IC) has the same error as GFS at forecast hour zero.
However, the two swapped models rapidly converge to the
same error as models that use consistent initial conditions during training and inference.
For 10m wind speed, this convergence happens in the first 6~hour
time step.
For 2m temperature, ML-GFS-Base(G+H~IC)
loses skill almost immediately as well, whereas Nested-EAGLE(GFS~IC)
takes slightly longer to converge.
The slower convergence appears to be mostly due to orography, since the swapped
models also swap static data; i.e., Nested-EAGLE(GFS~IC) uses GFS orography over
CONUS.
Thus, the remaining differences between Nested-EAGLE and
Nested-EAGLE(GFS~IC) correspond to a temperature bias consistent with a lapse rate
acting on the elevation differences between GFS and HRRR
(see \cref{si:t2m-dz-attribution} for more details).

\begin{figure}[t]
    \centering
    \includegraphics[width=\textwidth]{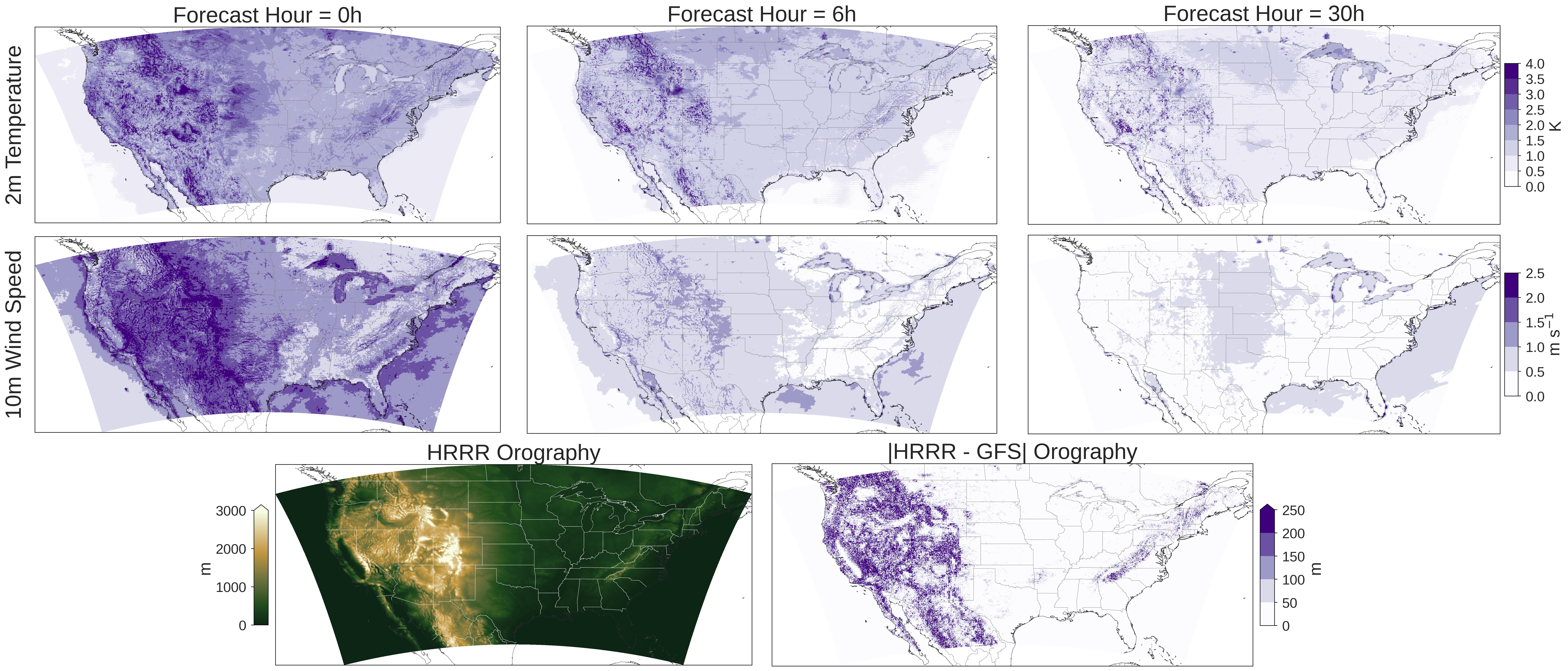}
    \caption{
        \textbf{
            Spatial RMSE between Nested-EAGLE and Nested-EAGLE(GFS~IC),
            where differences are restricted to the HRRR subdomain for clarity.
        }
        The figure shows the spatial distribution of errors between the
        green and purple dashed lines in \cref{fig:ic_experiment}.
        At forecast hour 0 (top rows, left), differences are entirely due to
        differences in GFS and HRRR analyses.
        After 6~hours (top rows, middle), the figure shows the difference between Nested-EAGLE with
        and without HRRR data over CONUS after a single forecast step.
        By 30~hours (top rows, right), the differences are negligible.
        The spatial RMSE represents the average over 293~forecasts.
        For a visual reference, the bottom row shows HRRR orography and the absolute value of
        differences between HRRR and GFS orography, regridded to 6~km
        resolution.
    }
    \label{fig:nested_swapics_spatial_rmse}
\end{figure}

Visualizing the spatial distribution of errors between Nested-EAGLE and
its swapped counterpart suggests that the model has learned to remove errors from the GFS
analysis.
\cref{fig:nested_swapics_spatial_rmse}
shows spatial maps of RMSE between Nested-EAGLE and Nested-EAGLE(GFS~IC),
averaged over the forecasts from the test period.
Within the first 6~hour time step, the model removes large errors from the Great Plains
and the coastlines.
By 30~hours into the forecast, the only notable differences between the two
experiments are in the 2m temperature field over regions with high elevation,
where persistent biases remain, owing to orographic differences noted earlier.
However, these differences are small relative to the initial error
between GFS and HRRR analysis.

\section{Precipitation Evaluation Over CONUS}
\label{sec:precip}

We evaluate Nested-EAGLE's ability to predict
6~hour accumulations of precipitation as a function of forecast hour
during the test period, February 2024--January 2025,
using a different protocol than in \cref{sec:prognostic_skill}.
Most importantly, we focus on the Fractions Skill Score
\citep[FSS;][]{roberts_scale-selective_2008} metric, which emphasizes both spatial
coherency and amplitude representation.
In essence, FSS gives better scores to precipitation forecasts
that ``look like'' observations; see \cref{subsec:precip-evaluation} for details.

We compare 1{,}426~forecasts from each model, initialized every 6~hours during the
test period.
We use more forecasts here than in \cref{sec:prognostic_skill} in order to
improve statistical power, since there are days without precipitation
events during the test year.
As a reference, we rely on the Analysis of Record for Calibration
\citep[AORC;][]{fall_aorc_2023} from NOAA's Office of Water Prediction (OWP); see
\cref{subsec:precip-evaluation} for more details.

\subsection{FSS Precipitation Skill: Amplitudes and Locations}
\label{subsec:fss_threshold}

\begin{figure}
    \centering
    \includegraphics[width=.9\textwidth]{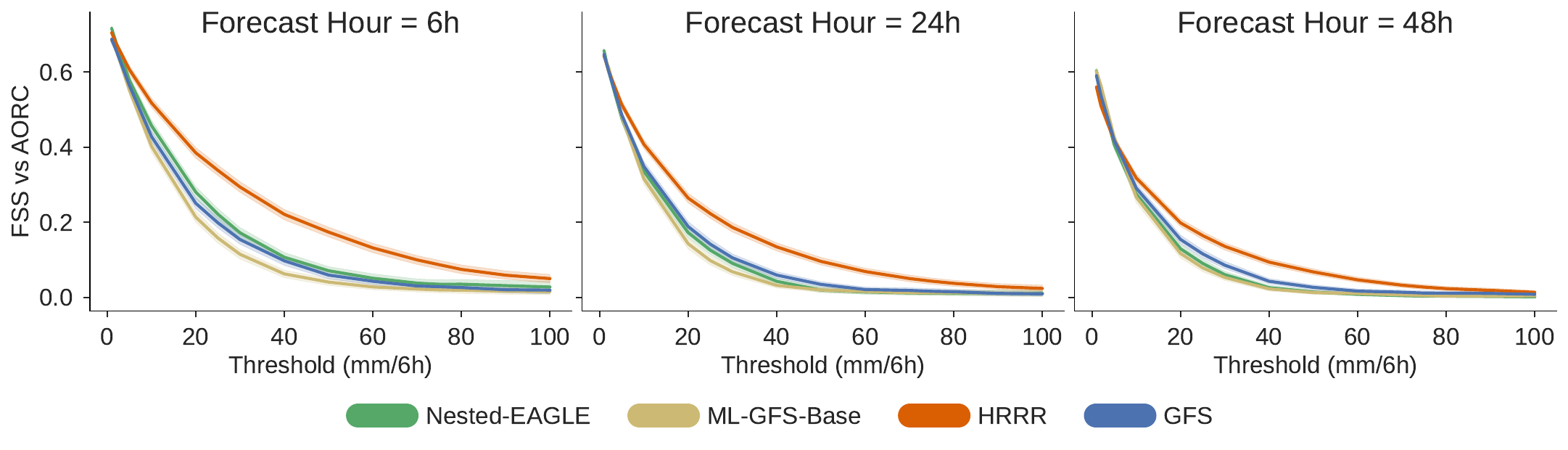}
    \caption{
        \textbf{
            FSS against AORC over CONUS during the test period,
            shown as a function of the absolute amplitude (mm/6h).
        }
        A value of 1 indicates a perfect FSS, i.e., higher is better.
        \meanci{1{,}426}
    }
    \label{fig:fss_threshold_conus}
\end{figure}

\cref{fig:fss_threshold_conus} quantifies the clearest limitation of
Nested-EAGLE: its precipitation forecasts are less skillful than HRRR, the
physics-based model it was trained on over CONUS.
Each panel of the figure shows FSS as a function of thresholds defined by
precipitation accumulated over the most recent 6~hours,
with panels indicating skill at 6, 24, and 48~hours of lead time.
HRRR is clearly the
most skillful forecast model at all threshold levels, except for the smallest
thresholds (1--2~mm/6h) where
Nested-EAGLE has a slight advantage.
At 6~hours of lead time, Nested-EAGLE is only marginally more skillful than
GFS despite operating at a higher resolution.
Similarly, ML-GFS-Base shows significantly lower skill than its physical
forecast model counterpart, GFS, and has the lowest skill of all models
evaluated.
All models lose skill with increasing lead time, and
the skill of the two ML models converges with increasing lead.
We surmise that this convergence occurs because, at longer leads, predictable
precipitation events are largely driven by forcing outside of the nested target
region, where the ML models have the same training data.

Nested-EAGLE and ML-GFS-Base produce less skillful precipitation forecasts relative to HRRR
and GFS, respectively, because they are trained deterministically with an MSE loss function.
More specifically, the MSE loss function suffers from the double-penalty effect
\citep{rossa_overview_2008},
leading to predictions that blur the small-scale features
\citep[as noted by many, e.g.,][]{lam_learning_2023}.
The relatively blurred small scales result in diminished local maxima, and
reduced skill at higher precipitation thresholds.

\subsection{FSS Precipitation Skill: Location Only}
\label{subsec:fss_percentile}

%\begin{wrapfigure}{r}{.62\textwidth}
\begin{figure}
    \centering
    \includegraphics[width=.9\textwidth]{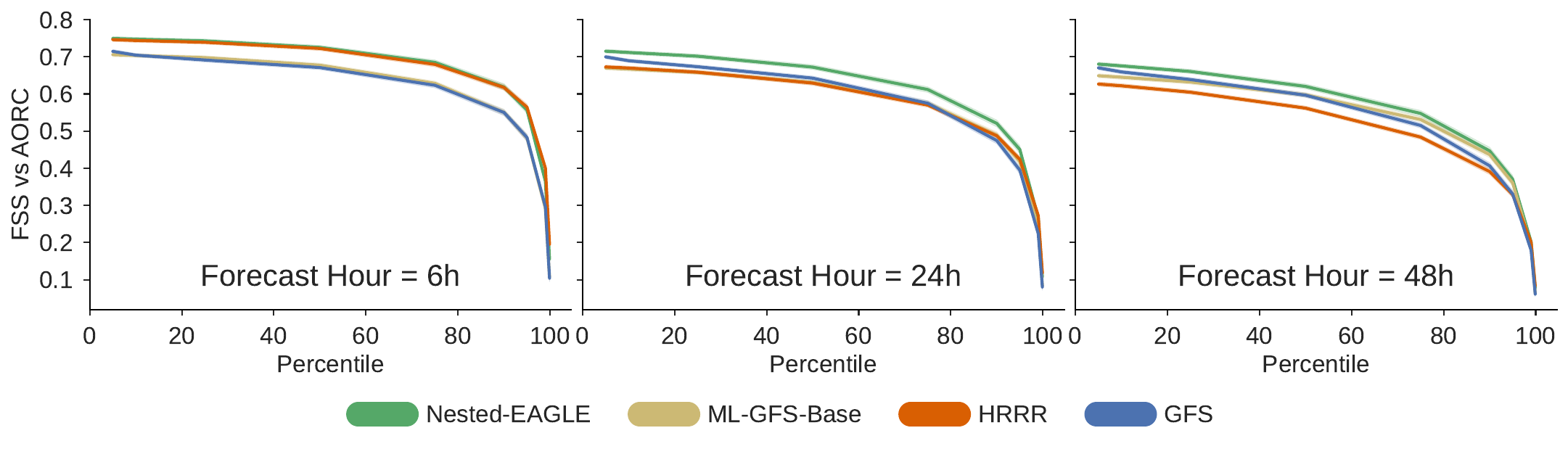}
    \caption{
        \textbf{
            FSS against AORC over CONUS during the test period,
            shown as a function of percentile represented by each model.
        }
        As in \cref{fig:fss_threshold_conus},
        a value of 1 indicates a perfect FSS, i.e., higher is better.
        \meanci{1{,}426}
    }
    \label{fig:fss_percentile_conus}
\end{figure}
%\end{wrapfigure}

Despite the blurring imposed by deterministic training with an MSE
loss,
\cref{fig:fss_percentile_conus} shows that Nested-EAGLE represents the location of
precipitation events well.
Each panel of the figure shows FSS as a function of percentile,
defined by the amplitude that each model is able to represent.
Viewing FSS from this perspective controls for each model's amplitude bias,
isolating each one's ability to capture the position of precipitation events
\citep{roberts_scale-selective_2008}.
See \cref{si:extended-precip} for a comparison of monthly
precipitation amounts during the validation period, which provides
qualitative corroboration that Nested-EAGLE places storms in the right location,
despite blurring the amplitudes.

At 6~hours of lead time, Nested-EAGLE and HRRR are the most skillful,
significantly outperforming ML-GFS-Base and GFS.
Additionally, at this lead time,
the ML models are indistinguishable from their physical model counterparts,
because the 0--6~hour forecasts are used as training targets.
Taken together, these results emphasize the importance of incorporating HRRR data via
the nesting procedure, since Nested-EAGLE is able to benefit from HRRR's
superior precipitation skill.

As in \cref{subsec:fss_threshold}, all models show decreasing FSS with
lead time, even in this percentile view.
However, the rankings change with lead.
Beyond 6~hours, Nested-EAGLE shows the highest skill scores, ML-GFS-Base
gradually outperforms GFS, and HRRR skill drops below GFS.
ML-GFS-Base skill approaches
Nested-EAGLE's with longer lead times, although this time the ML model skill
converges slower than in \cref{subsec:fss_threshold}.

We attribute HRRR's reduced skill, at least when viewed in this percentile
framing, to the challenging nature of capturing the
location of lower-likelihood events at longer lead times with deterministic,
high-resolution physical forecast systems.
As lead time extends, HRRR departs from the constraints
provided by data assimilation and tends toward the numerical model's attractor,
which is distinctly different from nature.
On the other hand, Nested-EAGLE is trained to produce HRRR's 0--6~hour
precipitation accumulation at all lead times, allowing the ML model's forecasts
to benefit from the data assimilation analysis further into the forecast.
ML-GFS-Base similarly outperforms GFS at longer leads,
consistent with MLWP models propagating the data assimilation constraints further
into the forecast than their traditional counterparts.

\section{Discussion}
\label{sec:discussion}

Operational centers provide forecast products to help inform the public on a
wide range of time scales, from hourly storm ``nowcasting'' to seasonal guidance.
While practicality plays a role, computational constraints have historically
necessitated developing independent systems for each forecast product.
We argue that collecting and synthesizing information from distinct sources
can be cumbersome for end users, and suggest that an integrated short- and
medium-range prediction system would be useful to the public.

With this goal in mind, we present Nested-EAGLE: a global 0.25$^\circ$ MLWP model
with a 6~km refinement over CONUS, trained on HRRR data nested inside of GFS
data.
Evaluating over CONUS against in situ observations shows that Nested-EAGLE has
significantly lower RMSE for near-surface quantities when compared to
NOAA's operational physics-based forecast systems, and ML-GFS-Base,
an MLWP baseline trained only on GFS data.
For example, Nested-EAGLE's 2m temperature forecasts over CONUS maintain a
significant lead over GFS for 7.5~days, and ML-GFS-Base for 6.5~days.
In the upper atmosphere and outside of CONUS, Nested-EAGLE produces skill
similar to
ML-GFS-Base, and in many cases these two MLWP models are comparable to GFS.
In summary, our results show that the biggest gains from the nesting
methodology are in near-surface fields over CONUS, stemming from
incorporating HRRR's improved representations of these quantities into the
model during training.
Nested-EAGLE is one of many exploratory MLWP models being tested at NOAA,
alongside its global \citep{sadeghi_tabas_gfs-powered_2025} and
LAM \citep{abdi_hrrrcast_2026} counterparts.
The model is currently producing forecasts for 2026 in a quasi-operational
environment, with an archive of reforecasts to be published soon.

Nested-EAGLE represents our current ability to bridge short- and medium-range
weather forecasting into one modeling system.
However, in its current form, the model is
deterministic rather than probabilistic,
and operates with a time step (6~hour) and
horizontal resolution (0.25$^\circ$/6~km)
that are too coarse to fully resolve the convective and storm scales that
underlie short-range weather.
Our FSS evaluation highlights how some of these limitations currently manifest.
That is, precipitation forecasts from Nested-EAGLE are less skillful and
under-represent high-to-extreme precipitation amounts compared to HRRR,
although they remain comparable to GFS.
Addressing these limitations will be necessary to capture a wider
range of spatiotemporal phenomena, but will also require reconciling some
trade-offs.
For example, reducing the time step size would allow Nested-EAGLE to capture
smaller-scale phenomena, but could introduce more error for 10--15~day forecasts
due to numerical instabilities.
Future developments could consider techniques like
temporal downscaling \citep{ingstad_hourglass_2026},
time-step-aware multilead training
\citep{nguyen_scaling_2023,abdi_hrrrcast_2026,weyn_aurora_2026},
or a mixture-of-experts approach
\citep{pathak_learning_2026,brenowitz_climate_2025}
to provide well-resolved short-range forecasts while remaining stable at longer
lead times.
Additionally, given the results by
\citet{lang_aifs-crps_2026,alet_skillful_2025,bonev_fourcastnet_2025},
we suggest that training a stochastic model with a continuous ranked probability
score loss function, perhaps with a spectral component as done by
\citet{nordhagen_high-resolution_2025},
would be fruitful for improving both phenomenological and uncertainty representation.
Subsequent work could focus separately on realism and uncertainty \citep{pereira_learning_2026}.

Another area of future work is to expand our training data.
In our prognostic evaluation, we found that, aside from near-surface variables
over CONUS, Nested-EAGLE performs comparably to a model trained only on GFS
data (ML-GFS-Base).
In contrast to the expectation that MSE-trained ML models should have lower RMSE than
their physical model counterparts,
these two ML models were largely comparable to GFS in terms of RMSE, except in
the low-level atmosphere and in the tropics (see extended evaluation in
\cref{si:extended-regional}).
We suggest that this discrepancy is due to the relatively
limited 8~year training datasets used in our work.
As a comparison, many other MLWP models are able to produce significantly lower
RMSE than physics-based models for atmospheric quantities like 500~hPa
geopotential height.
For example, \citet{lang_aifs_2024} show significant gains over the IFS after
training on 40~years of ERA5 reanalysis data, and then fine-tuning on 2~years of
operational IFS analysis data.
An obvious next step for Nested-EAGLE is to incorporate a pretraining schedule
similar to \citet{nipen_regional_2026}, and train first on ERA5 before using the
GFS and HRRR archives.
This work is already in active development.

More generally though, we suggest that future work should incorporate
observations directly into the model development.
Our analysis of precipitation event position provides some motivation along
these lines.
We showed that Nested-EAGLE accurately predicts precipitation
location, despite blurring the amplitudes.
This result is analogous to previous work showing that MLWP models tend to
accurately predict hurricane track, but fail to adequately forecast hurricane
intensity \citep[e.g.,][]{demaria_operations-based_2025}.
In our case, we suggest that Nested-EAGLE's advantage is due to the fact that it is
trained on 0--6~hour HRRR forecasts, which are still highly constrained by data
assimilation.
While our results show that Nested-EAGLE propagates this skill for longer lead
times than current operational models,
it cannot surpass HRRR's forecast skill at 6~hours in its current form with a
deterministic 6~hour time step, since this is its training target.
Thus, MLWP models could improve by incorporating observations directly into the learning
objective.
With this in mind, recent work by \citet{miralles_pointwise_2026} discusses some of the benefits
and challenges associated with learning from heterogeneous observation sources.

Finally, our experiments with perturbed, or ``swapped'', initial conditions not
only motivate learning from observations, but also motivate developing MLWP
models that operate on observational inputs.
More specifically, we showed that Nested-EAGLE produces comparable skill whether
it is initialized with GFS initial conditions only, or with the typical HRRR
and GFS combination.
Our spatial evaluation indicates that the model acts as a sort of bias
corrector, rapidly removing errors along the Great Plains and coastlines
from the GFS initial conditions.
Collectively, this evaluation confirms that, at least for the first
$\sim$30~hours of forecast lead time,
Nested-EAGLE's skill improvements are not due to better initial conditions at inference time,
but are instead learned from the HRRR data during training.
On the one hand, the results provide confidence that Nested-EAGLE can be robust
to degraded initial conditions, perhaps due to missing observational constraints
during the data assimilation phase.
However, these experiments also imply that improved initial conditions, for
example due to a system upgrade or increased observations, would not immediately
improve the model at forecast time.
Instead, improved initial conditions would need to be
incorporated into training for the ML model to realize their benefits.
While adding observational constraints to the loss
function could sidestep the need to nest or otherwise combine analyses,
our results suggest that retraining would still be necessary.
Thus, future work should consider modeling frameworks that
explicitly operate on observational inputs in order to respond more directly to
improved observational constraints.
Recent advancements in ``end-to-end'' ML forecasting
\citep{pinnington_aifs-dop_2026,pathak_learning_2026,zhao_skillful_2026,allen_end--end_2025,wang_xichen_2025,sun_data--forecast_2025}
and ML state estimation \citep{gupta_healda_2026} offer promise in this
direction.

\section{Methods}
\label{sec:methods}

\subsection{Nested-EAGLE Model Design}
\label{subsec:nested-eagle-design}

Nested-EAGLE is an autoregressive model parameterized by a deep neural network,
mapping the weather state $\state$ at times $t$ and $t-6~\text{hours}$ to
a prediction $\pred$ at $t+6~\text{hours}$.
That is, the model defines the mapping
\begin{equation}
    \model_{\params}\left(\state(t), \state(t-6\text{h})\right) = \pred(t+6\text{h}) \, .
    \label{eq:model}
\end{equation}
\cref{table:variables} lists the set of variables that define the model's state space.
Data for all variables except accumulated precipitation come from GFS and HRRR
forecast hour 0 output, which is essentially analysis data.
For each timestamp $t$, precipitation is taken as the 6~hour
accumulation forecast initialized 6~hours prior.
See \cref{si:sample-forecasts} for snapshots of several variables from a sample
forecast.

We used 8~years of data for training (February 2015--January 2023),
1~year for validation (February 2023--January 2024),
and 1~year for testing (February 2024--January 2025).
We built our model using the \texttt{anemoi} framework \citep[see][]{lang_aifs_2024},
and developed separate Python packages, \texttt{ufs2arco}
\citep{smith_ufs2arco_2026} and \texttt{eagle-tools} \citep{smith_eagle-tools_2026}, to
prepare datasets and perform evaluation, respectively.

\begin{table}[t]
\centering
\renewcommand{\arraystretch}{1.2}
\setlength{\tabcolsep}{4pt}
\begin{tabular}{@{}l@{\hspace{1em}}l@{}}
% --- LEFT: model outputs ---
\begin{tabular}[t]{@{}l@{\hspace{1em}}l@{}}
\hline
\multicolumn{2}{@{}l}{\textbf{Prognostic Fields} (Inputs \& Outputs)}\\
\textit{3D Atmospheric} & \textit{2D}\\
\hline
Geopotential Height & Surface Pressure\\
Zonal Wind          & 10m Zonal Wind\\
Meridional Wind     & 10m Meridional Wind\\
Vertical Velocity   & Surface Temperature\\
Temperature         & 2m Temperature\\
Specific Humidity   & 2m Specific Humidity\\
\hline
\end{tabular}
&
% --- RIGHT: forcing inputs + diagnostics ---
\begin{tabular}[t]{@{}l@{\hspace{1em}}l@{}}
\hline
\multicolumn{2}{@{}l}{\textbf{Forcing Fields} (Inputs Only)}\\
\textit{Time Varying} & \textit{Static}\\
\hline
Solar Insolation             & Land-Sea Mask\\
Cosine \& Sine of Latitude   & Orography\\
Cosine \& Sine of Longitude  & \\
Cosine \& Sine of Julian Day & \\
Cosine \& Sine of Local Time & \\
\hline
\multicolumn{2}{@{}l}{\textbf{Diagnostic Fields} (Outputs Only)}\\
\hline
6~hour Accumulated Precipitation & 80m Zonal Wind\\
80m Meridional Wind         & \\
\hline
\end{tabular}
\end{tabular}
\caption{
    \textbf{State space definition.} The 3D atmospheric fields are represented on
    12 pressure levels: 100, 150, 200, 250, 300, 400, 500, 600, 700, 850, 925, and
    1000~hPa.
    All fields are normalized by their Z-score ($(x-\mu)/\sigma$) with the
    following exceptions.
    Specific Humidity, 2m Specific Humidity, and Accumulated Precipitation
    are normalized by the standard deviation ($x/\sigma$).
    Orography is normalized by its maximum value ($x/\max(x)$),
    and all other forcing fields are not normalized.
    Specific Humidity, 2m Specific Humidity, and Accumulated Precipitation are
    bounded to positive values with $\text{ReLU}(x)=\max(x,0)$, following
    \citet{moldovan_aifs_2026}.
}
\label{table:variables}
\end{table}

The neural network that defines Nested-EAGLE consists of encoder, processor, and
decoder modules, similar to many MLWP models
\citep[e.g.,][]{lam_learning_2023,lang_aifs_2024,nipen_regional_2026}.
We describe the key differentiating elements here and provide more model details in
\cref{si:model-dev}.

The encoder maps the weather state from the data space to the latent mesh,
using the same graph-transformer design as \citet{lang_aifs_2024}.
The decoder uses the same graph-transformer blocks, and performs the reverse
mapping: from the latent mesh to the data space.
The Nested-EAGLE data space is defined by joining
GFS data, archived on a 0.25$^\circ$ latitude-longitude grid,
with HRRR data on a Lambert conformal conic projection over CONUS.
For the current version of Nested-EAGLE,
HRRR data are conservatively regridded from 3~km to 6~km resolution.
We do not retain GFS data nodes over the HRRR subdomain (i.e., over CONUS),
similar to how
\citet{nipen_regional_2026,bano-medina_regional_2025}
define nested, or ``stretched'', grids.
In our work, we design the latent mesh in a similar fashion to how the data
space is created, except at a resolution that is approximately 16 times coarser
(i.e., factor of $\sim$4 in latitude and longitude).
We combine a global octahedral reduced Gaussian grid at O96 resolution
(approximately 1$^\circ$, or $\sim$100~km resolution, with 40{,}320 nodes)
together with a subset of the HRRR grid, coarsened to 24~km resolution.

\begin{figure}[t]
    \centering
    \includegraphics[width=.8\textwidth]{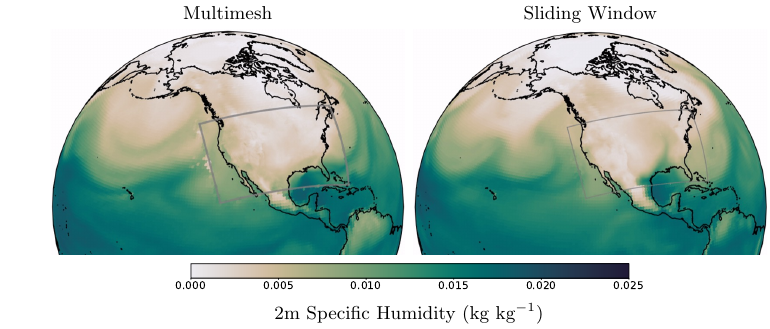}
    \caption{
        \textbf{10~day forecasts after initialization at 0000 UTC 9 March 2023 from
        prototypes using different latent meshes and processor architectures.}
        The left panel comes from a model with a graph-transformer
        processor acting on an icosahedral
        multimesh with $\sim$158~km and $\sim$20~km resolutions globally and over
        CONUS, respectively.
        The right panel comes from a model with a sliding-window transformer
        processor and a latent mesh with 2$^\circ$ and 30~km resolutions
        globally and over CONUS, respectively.
        Note that both forecasts stem from earlier prototypes trained on GFS
        data regridded to 1$^\circ$ and HRRR data regridded to 15~km resolutions.
    }
    \label{fig:processor-artifacts}
\end{figure}

Our latent mesh and processor design differs from
\citet{nipen_regional_2026,bano-medina_regional_2025}, who use an icosahedral multimesh
latent space with triangular elements, employing
higher refinements over their target regions, together with a
graph-transformer processor.
Our earliest prototypes used the same design, but suffered from artifacts at the
boundary between the global and target subdomains.
We therefore designed the custom latent mesh noted earlier in order to use a sliding-window
transformer processor, as in \citet{lang_aifs_2024}, hypothesizing that a more
seamless numerical stencil would help remove the artifacts.
\cref{fig:processor-artifacts} shows
10~day forecasts from two early prototypes, confirming that
the sliding-window transformer removed the grid artifacts.

Nested-EAGLE does require a relatively large ``window size''
in order to adequately capture enough latent nodes in a single processor stage;
see \cref{si:window-size} for sensitivity experiments related to this
hyperparameter and \cref{fig:nested_illustration} for illustrative intuition.
However, the sliding-window transformer still reduced computational costs by
$\sim$33\% relative to the multimesh graph-transformer processor.
We attribute the efficiency gains to several factors.
First, total runtime is relatively insensitive to
some extra compute costs, since the I/O and memory operations stemming from the
four states in the model's inputs
(two from HRRR, two from GFS) tend to dominate.
Second, the sliding-window attention acts on consecutive slices of the latent
state arrays, whereas pure graph-based approaches rely on random access
patterns that often result in cache misses and suboptimal GPU utilization.
Finally, we attribute computational speed to the highly engineered
FlashAttention implementation by
\citet{dao_flashattention_2022,dao_flashattention-2_2023}, which is used in the
sliding-window processor.

The other most notable design difference between our work and
\citet{nipen_regional_2026} is that we do not pretrain our model using ERA5.
More specifically, \citet{nipen_regional_2026} opted to use only 2.5~years of IFS and MEPS
archives for training data, in order to avoid inconsistencies between different
MEPS forecast model versions.
On the other hand, we opted to use all HRRR and GFS data available since
February 2015, the earliest publicly available GFS data at 0.25$^\circ$ to our knowledge
\citep{national_centers_for_environmental_prediction_national_weather_service_noaa_us_department_of_commerce_ncep_2015,national_centers_for_environmental_prediction_national_weather_service_noaa_us_department_of_commerce_ncep_2015-1}.
With 8~years of available training data, our initial prototypes were able to
produce lower RMSE than GFS and HRRR.
After switching to a sliding-window transformer architecture, we did not observe
any obvious signs of having insufficient data,
and therefore neglected the pretraining step.
\cref{table:training} shows our full training schedule.
However, work is already ongoing to incorporate pretraining and further improve skill,
especially outside of the HRRR subdomain.

\begin{table}[t]
\centering
\renewcommand{\arraystretch}{1.2}
\begin{tabular}{@{}l c c c@{}}
\hline
\textbf{Hyperparameter} & \textbf{Stage A} & \textbf{Stage B} & \textbf{Stage C}\\
\hline
Optimization Iterations        & 60{,}000 & 5{,}760 & 5{,}760\\
Epochs        & $\sim$82 & 8 & 8\\
Linear Warmup Steps            & 1{,}000  & 100     & 100\\
Maximum Effective Learning Rate & $1.0\times10^{-3}$ & $1.0\times10^{-4}$ & $1.0\times10^{-5}$\\
Minimum Learning Rate          & $3.0\times10^{-7}$ & $3.0\times10^{-7}$ & $3.0\times10^{-7}$\\
Autoregressive Rollout Steps   & 1 & 2 & 2--4\\
\hline
\end{tabular}
\caption{
    \textbf{Training Schedule}.
    All stages used a batch size of 16, and our 8~year training dataset has
    11{,}616 samples, resulting in about 726 optimization iterations
    (or batches) per epoch in Stage A.
    We used an AdamW optimizer \citep{loshchilov_decoupled_2018} with parameters $\beta_1=0.9,
    \beta_2=0.95$.
    The loss is a weighted MSE function, where the loss over the
    target subregion is upweighted to 10\% of the total loss; see
    \cref{si:lam-loss} for details on this choice.
    Training required approximately 45~hours of wall-clock time on the National Energy Research Scientific
    Computing Center's (NERSC) Perlmutter machine, using 16~nodes each with 4~A100s and
    40~GB of GPU RAM per A100.
    We distributed the model across a single node, i.e., 4~GPUs per model
    instance.
}
\label{table:training}
\end{table}

\subsection{ML-GFS-Base Model Design}
\label{subsec:baseline-design}

We compare Nested-EAGLE to ML-GFS-Base,
a baseline single-resolution MLWP model trained solely on GFS data,
in order to isolate the impacts of incorporating HRRR data through the
nesting process.
As such, we designed this baseline model to be as close a match to
Nested-EAGLE as possible.
The only design differences, aside from the datasets, are as follows.
The ML-GFS-Base data space is a uniform 0.25$^\circ$ latitude-longitude grid, matching
Nested-EAGLE nodes outside of the HRRR subdomain.
The latent mesh is a single-resolution O96 octahedral reduced Gaussian, again
matching Nested-EAGLE outside of the HRRR subdomain.
The sliding-window transformer processor uses a window size of 2{,}250, which is
much smaller than 8{,}168 used in Nested-EAGLE owing to the fact that there are
far fewer nodes in the latent mesh.
Finally, the ML-GFS-Base loss function normalizes the loss at each node
based on its grid-cell area estimated by a Voronoi tessellation, and there is no
renormalization after this.
On the other hand, the loss for Nested-EAGLE is reweighted after this grid-cell
area normalization so that the HRRR subdomain accounts for 10\% of the loss fraction
(see \cref{si:lam-loss} for sensitivity test results).

\subsection{Forecast Model Baselines}
\label{subsec:baseline-datasets}

We compare Nested-EAGLE to the two operational numerical weather prediction
forecast models that underlie
the training data: NOAA's HRRR and GFS.
The HRRR is a regional model covering CONUS, operating at 3~km spatial resolution, and is
initialized every hour from NOAA's 13~km resolution RAP \citep{dowell_high-resolution_2022}.
The GFS is NOAA's central deterministic medium-range weather forecast system,
initialized four times daily at 6~hour intervals by GDAS.
Our validation and test periods cover February 2023--January 2024 and February
2024--January 2025, respectively.
Thus, we compare Nested-EAGLE to HRRR version 4 and
GFS version 16, the current operational forecast systems at the time of writing.

\subsection{Prognostic Evaluation}
\label{subsec:prognostic-evaluation}

\begin{figure}[t]
    \centering
    \includegraphics[width=.7\textwidth]{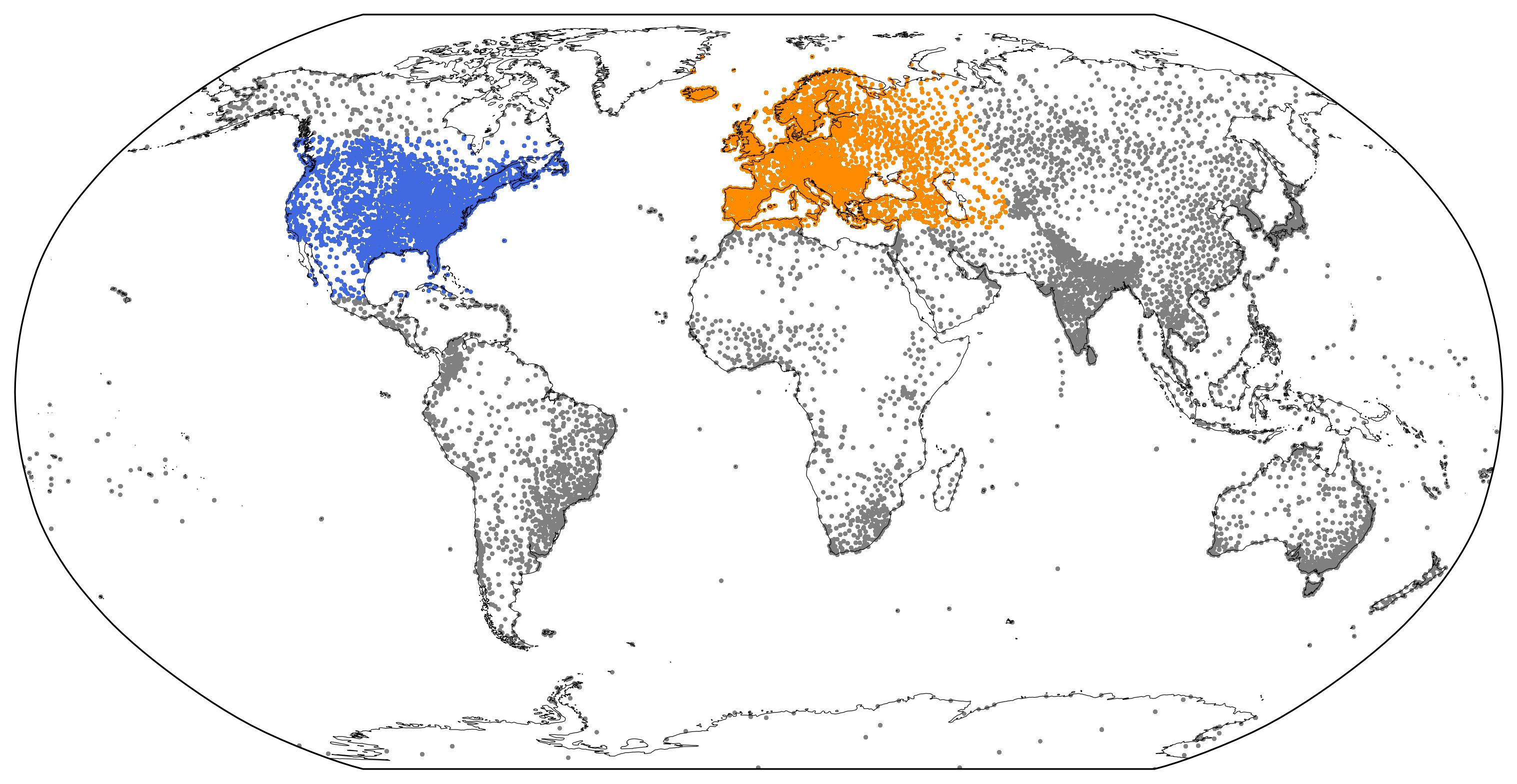}
    \caption{
        \textbf{24~hours of conventional observational coverage on 7 December 2024.}
        Observations over the U.S. and Europe are colored in blue and orange,
        respectively, as these subregions are discussed in
        \cref{subsec:conus_skill,subsec:europe_skill}.
        Observations everywhere else are indicated by gray dots.
    }
    \label{fig:obs-coverage}
\end{figure}

We evaluate forecasts from each model using
in situ (conventional) observations as a reference.
The observations come from the NOAA NASA Joint
Archive \citep[NNJA;][]{nnja_obs_data},
which hosts a collection of data sources, e.g.,
from dropsondes, rawinsondes, aircraft, and stations.
We access the data using Brightband's
Python API \citep{hans_mohrmann_2025_16749100}.
\cref{fig:obs-coverage} shows a representative snapshot of observational
coverage during one day in the test period,
highlighting the subsets over CONUS and Europe discussed in
\cref{subsec:conus_skill,subsec:europe_skill}, respectively.

We evaluate the forecasts in terms of their RMSE
against the observations noted above.
For a given valid time $t$, we select observations that are within one hour
($t\pm30$~minutes) to get $\obs(t)$, and bilinearly interpolate the predicted
state from each model to these observation locations to get $\opred(t)$.
In our evaluation framework, the number of observations and their locations
change with $t$.
However, to streamline the presentation here, we assume that there are $N_o$
observations at each time $t$, indexed by $i\in\{1, 2, ..., N_o\}$.
For each forecast initialization, indexed by $j\in\{1, 2, ..., N_f\}$, we compute a
single error value at each forecast hour $\Delta t$ as
\begin{equation}
    \text{RMSE}_j(\Delta t) = \sqrt{
        \dfrac{1}{N_o}
        \sum_{i=1}^{N_o}
        \left(\opred_{ij}(\Delta
        t)-\obs_{i}(t_j + \Delta t)\right)^2
    } \, ,
    \label{eq:rmse}
\end{equation}
and report the median over the $N_f$ initializations.
We aggregate RMSE using the median rather than the mean when comparing against
in situ observations because the confidence intervals on mean RMSE are too large
to discern differences for 2m specific humidity.
We suspect that outliers obfuscate the estimate of mean RMSE for specific
humidity.
Using either median RMSE or mean absolute error tends to show a clear picture
for this quantity, and other variables behave consistently no matter what, so
we use the former here.
We evaluate the RMSE at 6~hour increments, up to a 15~day lead time.

We estimate confidence intervals for our aggregated error estimates using the Python
package \texttt{seaborn} \citep{waskom_seaborn_2021}.
For completeness, we state the general procedure here, which we use for other
metrics like FSS; see \cref{subsec:precip-evaluation}.
For all confidence intervals, we use a nonparametric bootstrapping methodology
over forecast initializations.
Let $e_j(\Delta t)$ denote a per-initialization skill value and $\mathcal{S}$ the
statistic used to aggregate it: $e_j = \text{RMSE}_j$ with
$\mathcal{S}=\text{median}$ for the prognostic fields compared against
observations, and $e_j = \text{FSS}_j$ with
$\mathcal{S}=\text{mean}$ for precipitation (see
\cref{subsec:precip-evaluation}).
For each of $N_B$ resamples, we draw initialization indices
$j^{(b)}_1, \dots, j^{(b)}_{N_f}$ uniformly with replacement from
$\{1, 2, ..., N_f\}$ and recompute
\begin{equation}
    e^{(b)}(\Delta t) = \mathcal{S}\left(
        \left\{ e_{j^{(b)}_k}(\Delta t) \right\}_{k=1}^{N_f}
    \right) \, ,
    \qquad b = 1, \dots, N_B \, .
    \label{eq:bootstrap}
\end{equation}
The shading in our figures spans the 2.5th and 97.5th percentiles of
$\{e^{(b)}(\Delta t)\}_{b=1}^{N_B}$, i.e., a percentile bootstrap 95\% confidence
interval, using $N_B=1{,}000$ resamples.
We consider a difference between two models to be statistically significant when
their intervals do not overlap, which is a conservative criterion given that all
models are verified against the same observations at the same initializations.

Conducting the evaluation against observations is critical,
because Nested-EAGLE combines analysis states from both GFS and HRRR.
Comparing against observations allows us to control for regions in the analyses
that differ from one another but are unconstrained by data.

\subsection{Precipitation Evaluation}
\label{subsec:precip-evaluation}

We compare the precipitation fields separately from the prognostic model
outputs.
Because it is challenging to find high-resolution, high-quality observational
estimates of precipitation across the globe, we limit our evaluation to the CONUS region,
using version 1.1 of NOAA OWP's AORC \citep{fall_aorc_2023} as a reference.
Notably, AORC provides an estimate of accumulated surface precipitation at an hourly frequency from
1979 through 2025 at $\sim$800~m resolution.
The precipitation estimate is derived from a variety of sources, relying largely
on CONUS-wide Stage IV radar data in more recent years
\citep{bytheway_representation_2026}.
To facilitate comparison, we conservatively regrid all datasets to the 6~km resolution of the
Nested-EAGLE grid over CONUS, such that the total precipitation amounts are preserved.
Finally, the hourly AORC data are aggregated so that all datasets represent
6~hour accumulations.

To evaluate precipitation forecast skill we use
the FSS metric \citep{roberts_scale-selective_2008}, which emphasizes both spatial
coherency and amplitude representation.
In the following definitions, we drop explicit dependence on forecast
initialization and lead time, and simply state that we compute FSS for each lead
time, averaged over all 1{,}426~initializations during the test period.
For a threshold $q$ and square box neighborhood $\mathcal{B}_r(m)$ of radius $r$
centered on grid cell $m$, the fractional coverage fields of the truth $\bm{p}$
and the prediction $\hat{\bm{p}}$ are
\begin{equation}
    \phi_m = \frac{\sum_{k \in \mathcal{B}_r(m)} v_k\, \mathds{1}_{p_k \ge q}}
                  {\sum_{k \in \mathcal{B}_r(m)} v_k},
    \qquad
    \hat{\phi}_m = \frac{\sum_{k \in \mathcal{B}_r(m)} v_k\, \mathds{1}_{\hat{p}_k \ge q}}
                        {\sum_{k \in \mathcal{B}_r(m)} v_k},
    \label{eq:fss_fraction}
\end{equation}
where $v_k \in \{0, 1\}$ flags valid grid cells. The FSS is then
\begin{equation}
    \mathrm{FSS}(q, r) = 1 -
        \frac{\sum_m \left(\hat{\phi}_m - \phi_m\right)^2}
             {\sum_m \left(\hat{\phi}_m^2 + \phi_m^2\right)},
    \label{eq:fss}
\end{equation}
where the sums run over all cells $m$ with at least one valid point in their
neighborhood.
In this work we use $r=25$~km, corresponding to evaluation ``boxes'' with
54~km sides, given the 6~km resolution.

We also employ a percentile-based variant of the score in order to control for
amplitude biases and highlight spatial positioning skill.
For this variant, the single
threshold $q$ is replaced by a distinct threshold for each field, set to that
field's own $P$-th percentile. Let $Q_P(\cdot)$ denote the $P$-th percentile
($P \in [0, 100]$) taken over the valid, wet grid cells across the full domain,
i.e., cells with $v_k = 1$ and a strictly positive value. The truth and
prediction thresholds are then
\begin{equation}
    q_{\bm{p}} = Q_P\!\left(\left\{\, p_k : v_k = 1,\ p_k > 0 \,\right\}\right),
    \qquad
    q_{\hat{\bm{p}}} = Q_P\!\left(\left\{\, \hat{p}_k : v_k = 1,\ \hat{p}_k > 0 \,\right\}\right).
    \label{eq:fss_percentile_threshold}
\end{equation}
The fractional coverage fields and score are computed exactly as in
\cref{eq:fss_fraction,eq:fss}, with the only change being that the exceedance
indicators $\mathds{1}_{p_k \ge q}$ and $\mathds{1}_{\hat{p}_k \ge q}$ are
replaced by $\mathds{1}_{p_k \ge q_{\bm{p}}}$ and
$\mathds{1}_{\hat{p}_k \ge q_{\hat{\bm{p}}}}$, respectively.
The percentile is taken over wet
($>0$) cells only because the large dry point-mass in precipitation would
otherwise force the low- and mid-percentile thresholds to $0$~mm, at which point
every cell satisfies the exceedance test and the score collapses to $\approx 1$;
the exceedance indicators themselves, however, still range over all valid cells.

We compute confidence intervals for FSS with the bootstrap in
\cref{eq:bootstrap}, taking $e_j$ as the FSS for initialization $j$ and
$\mathcal{S}$ as the mean over the 1{,}426~initializations.

\section{Data Availability}
\label{sec:data-availability}

The GFS archives used for training and as a baseline forecast model are
openly available from the National Center for Atmospheric Research (NCAR)
Research Data Archive for forecast initializations spanning February 2015--August 2026
\citep{national_centers_for_environmental_prediction_national_weather_service_noaa_us_department_of_commerce_ncep_2015,national_centers_for_environmental_prediction_national_weather_service_noaa_us_department_of_commerce_ncep_2015-1}.
In our work, for forecasts initialized January 2021 onward,
we used the GFS data made available by
NOAA's Open Data Dissemination (NODD) program,
accessed via the Amazon Web Services (AWS)
Registry of Open Data at
\url{https://registry.opendata.aws/noaa-gfs-bdp-pds}, accessed August 2025.
The HRRR archives used for training and as a baseline forecast model are made
available by the NODD program, and we accessed the data via the AWS Registry of
Open Data at
\url{https://registry.opendata.aws/noaa-hrrr-pds}, accessed August 2025.
The conventional observations used to evaluate the models were part of NNJA,
accessed December 2025
\citep{nnja_obs_data}.
In our work, we made extensive use of the AI-Ready version of the observational
data, available via the Brightband Python API
\citep{hans_mohrmann_2025_16749100}
and parquet-formatted data
available on Google Cloud Storage
\url{https://console.cloud.google.com/storage/browser/nnja-ai}.
The AORC dataset used as a reference for precipitation evaluation is made
available by the NODD program, and we accessed the data via the AWS Registry of
Open Data at
\url{https://registry.opendata.aws/noaa-nws-aorc}, accessed August 2025
\citep{fall_aorc_2023}.

\section{Code Availability}
\label{sec:code-availability}

All evaluation scripts and configuration files, specifying data ingest, model
design, training regimen, etc., can be found at
\url{https://github.com/NOAA-PSL/nested-eagle}.
See also a release from \citet{epic_nested_2026}, which details related
available resources, including neural network weights for Nested-EAGLE at
\url{https://eaglecheckpoints.blob.core.windows.net/eagle-checkpoints/nested-eagle/inference-last.ckpt}.
The repository makes use of several open-source packages.
We developed and used \texttt{ufs2arco} for data preprocessing and ingest
\citep{smith_ufs2arco_2026}.
We used the \texttt{anemoi} infrastructure for model development, training, and inference
available at
\url{https://github.com/ecmwf/anemoi-core} and
\url{https://github.com/ecmwf/anemoi-inference}
\citep[see][]{lang_aifs_2024}.
We developed and used the Python package \texttt{eagle-tools} for model evaluation
\citep{smith_eagle-tools_2026}.
Earlier developments also made use of the Python package \texttt{wxvx} for model evaluation, available at
\url{https://github.com/NOAA-GSL/wxvx}.

\section{Acknowledgements}
\label{sec:acknowledgements}

This research used resources of the National Energy Research Scientific
Computing Center (NERSC), a Department of Energy User Facility, under
Contract DE-AC02-05CH11231 using NERSC Award
GenAI@NERSC DDR-ERCAP0034078.
Funding for Basarab, Abdi, and Madden was provided by NOAA Cooperative Agreement
NA22OAR4320151 for the Cooperative Institute for Earth
System Research and Data Science (CIESRDS).
Smith's contributions to this work were primarily carried out at NOAA's
Physical Sciences Laboratory;
part of the evaluation and the preparation of this manuscript were completed
at the Nansen Environmental and Remote Sensing Center in Bergen, Norway.
We thank Niraj Agarwal, Monte Flora, Haonan Chen, Laura Slivinski, Chong-Chi Tong,
Bo Huang, and the members of NOAA's AI4NWP Forum for useful discussions related to the
work presented here.
We are grateful to the Bris team at Met Norway and the ECMWF \texttt{anemoi} developers
for making their model advancements available in an open source ecosystem.
The statements, findings, conclusions, and recommendations are those of the
author(s) and do not necessarily reflect the views of NOAA or the U.S.
Department of Commerce.

\bibliography{references}

\clearpage
\phantomsection
\pdfbookmark[1]{Supporting Information}{si-title}
\setsititle
\maketitle
% TOC of the supplement only. \startcontents begins recording a named
% ("si") contents list here -- after the main text -- so main-text
% sections are excluded; \printcontents typesets that list. Main-text
% PDF bookmarks are untouched.
\startcontents[si]
\printcontents[si]{}{1}{}
\clearpage

% ================================================================
%  Supporting Information numbering
% ----------------------------------------------------------------
%  \input this at the top of the SI content so that figures, tables,
%  equations, and sections are prefixed with "S" (S1, S2, ...) and
%  numbered independently of the main text. This works both when the
%  SI is appended to the preprint and when it is compiled standalone.
% ================================================================
\setcounter{figure}{0}   \renewcommand{\thefigure}{S\arabic{figure}}
\setcounter{table}{0}    \renewcommand{\thetable}{S\arabic{table}}
\setcounter{equation}{0} \renewcommand{\theequation}{S\arabic{equation}}
\setcounter{section}{0}  \renewcommand{\thesection}{S\arabic{section}}
\setcounter{tocdepth}{2} % 2 = subsections

% Resetting the counters above makes the SI numbers (1, 2, ...) collide
% with the main text's hyperref anchors (section.1, figure.1, ...), so
% links point at the wrong target. Give the SI its own unique anchors by
% redefining the hyper-counter macros (harmless when hyperref is absent).
\renewcommand{\theHsection}{S\arabic{section}}
\renewcommand{\theHsubsection}{S\arabic{section}.\arabic{subsection}}
\renewcommand{\theHfigure}{S\arabic{figure}}
\renewcommand{\theHtable}{S\arabic{table}}
\renewcommand{\theHequation}{S\arabic{equation}}

\section{Extended Prognostic Evaluation}
\subsection{Lead Time Gap}
\label{si:lead-time-equivalence}

\cref{sitab:equivalent-surface,sitab:equivalent-upper-air}
explicitly show the lead time gap, $\tau_\text{gap}$, described in
\cref{subsec:conus_skill,eq:skill-gap}.
The tables show the gap for each baseline model relative to Nested-EAGLE, based
on the
median Root-Mean-Squared Error (RMSE) from 293~forecasts as described in the main text.
As noted in \cref{subsec:prognostic-evaluation},
we consider a gap to be statistically significant if the 95\% confidence
interval from each model does not overlap.
We report statistically insignificant values in gray.

\begin{table}[H]
\centering
\footnotesize
\setlength{\tabcolsep}{5pt}
\renewcommand{\arraystretch}{1.05}
\begin{tabular}{@{}l|ccc|ccc|ccc@{}}
\hline
 & \multicolumn{3}{|c}{\textbf{10\,m Wind Speed}} & \multicolumn{3}{|c}{\textbf{2\,m Temperature}} & \multicolumn{3}{|c}{\textbf{2\,m Specific Humidity}}\\
\textbf{Lead Time} & \textbf{ML-GFS-Base} & \textbf{HRRR} & \textbf{GFS} & \textbf{ML-GFS-Base} & \textbf{HRRR} & \textbf{GFS} & \textbf{ML-GFS-Base} & \textbf{HRRR} & \textbf{GFS}\\
\hline
0\,h & $+84$ & \nsig{$0$} & $+84$ & $+84$ & \nsig{$0$} & $+84$ & $+84$ & \nsig{$0$} & $+84$\\
12\,h & $+66$ & $+54$ & $+84$ & $+72$ & $+42$ & $+84$ & $+78$ & $+54$ & $+84$\\
24\,h & $+54$ & $+60$ & $+78$ & $+66$ & $+48$ & $+78$ & \nsig{$+42$} & $+42$ & $+60$\\
\hline
36\,h & $+48$ & $+66$ & $+84$ & $+54$ & $+48$ & $+72$ & \nsig{$+54$} & $+60$ & $+66$\\
48\,h & $+42$ & $+66$ & $+72$ & $+48$ & $+48$ & $+60$ & \nsig{$+42$} & $+54$ & $+60$\\
\hline
60\,h & $+36$ & -- & $+72$ & $+42$ & -- & $+60$ & \nsig{$+42$} & -- & \nsig{$+60$}\\
72\,h & $+30$ & -- & $+60$ & $+36$ & -- & $+48$ & \nsig{$+30$} & -- & $+48$\\
\hline
84\,h & $+30$ & -- & $+60$ & $+30$ & -- & $+48$ & \nsig{$+6$} & -- & \nsig{$+36$}\\
96\,h & $+30$ & -- & $+66$ & $+24$ & -- & $+42$ & \nsig{$+24$} & -- & $+42$\\
\hline
108\,h & $+24$ & -- & $+54$ & $+24$ & -- & $+42$ & \nsig{$+18$} & -- & \nsig{$+30$}\\
120\,h & $+24$ & -- & $+48$ & $+24$ & -- & $+30$ & \nsig{$+6$} & -- & \nsig{$+24$}\\
\hline
132\,h & $+30$ & -- & $+54$ & $+24$ & -- & $+24$ & \nsig{$+12$} & -- & \nsig{$+30$}\\
144\,h & $+24$ & -- & $+60$ & $+12$ & -- & $+18$ & \nsig{$+6$} & -- & \nsig{$+18$}\\
\hline
156\,h & $+30$ & -- & $+54$ & $+12$ & -- & $+12$ & \nsig{$+6$} & -- & \nsig{$+12$}\\
168\,h & \nsig{$+24$} & -- & $+66$ & \nsig{$+18$} & -- & $+18$ & \nsig{$+18$} & -- & \nsig{$+12$}\\
\hline
180\,h & \nsig{$+24$} & -- & $+78$ & \nsig{$+12$} & -- & $+24$ & \nsig{$+6$} & -- & \nsig{$+12$}\\
192\,h & \nsig{$+36$} & -- & $+126$ & \nsig{$+12$} & -- & \nsig{$+12$} & \nsig{$0$} & -- & \nsig{$+12$}\\
\hline
204\,h & \nsig{$+54$} & -- & $>+156$ & \nsig{$+18$} & -- & \nsig{$+6$} & \nsig{$0$} & -- & \nsig{$+12$}\\
216\,h & \nsig{$+102$} & -- & $>+144$ & \nsig{$+12$} & -- & \nsig{$+12$} & \nsig{$0$} & -- & \nsig{$+12$}\\
\hline
228\,h & \nsig{$+90$} & -- & $>+132$ & \nsig{$+6$} & -- & \nsig{$+12$} & \nsig{$+18$} & -- & \nsig{$+18$}\\
240\,h & \nsig{$+78$} & -- & $>+120$ & \nsig{$0$} & -- & \nsig{$+18$} & \nsig{$-12$} & -- & \nsig{$+42$}\\
\hline
\end{tabular}
\caption{
    \textbf{Lead time gain for the near-surface fields.}
    Each value indicates the lead time gain ($\tau_\text{gap}$, see
    \cref{eq:skill-gap}) that Nested-EAGLE has for a given model and variable.
    Positive values indicate the additional lead time that Nested-EAGLE has
    lower RMSE than the listed model.
    Gray values indicate statistically insignificant differences between the
    median RMSE, based on the 293~forecasts described in the main text and a
    95\% confidence interval.
}
\label{sitab:equivalent-surface}
\end{table}

\begin{table}[H]
\centering
\footnotesize
\setlength{\tabcolsep}{5pt}
\renewcommand{\arraystretch}{1.05}
\begin{tabular}{@{}l|ccc|ccc|ccc@{}}
\hline
 & \multicolumn{3}{|c}{\textbf{850\,hPa Temperature}} & \multicolumn{3}{|c}{\textbf{500\,hPa Geopotential Height}} & \multicolumn{3}{|c}{\textbf{250\,hPa Wind Speed}}\\
\textbf{Lead Time} & \textbf{ML-GFS-Base} & \textbf{HRRR} & \textbf{GFS} & \textbf{ML-GFS-Base} & \textbf{HRRR} & \textbf{GFS} & \textbf{ML-GFS-Base} & \textbf{HRRR} & \textbf{GFS}\\
\hline
0\,h & $+6$ & \nsig{$0$} & $+6$ & $+18$ & \nsig{$0$} & $+18$ & $+6$ & \nsig{$0$} & $+6$\\
12\,h & \nsig{$0$} & \nsig{$0$} & \nsig{$+12$} & $+12$ & $+12$ & $+18$ & \nsig{$0$} & \nsig{$+6$} & \nsig{$-6$}\\
24\,h & \nsig{$+6$} & $+18$ & $+24$ & $+6$ & $+12$ & $+12$ & \nsig{$-6$} & $+12$ & \nsig{$0$}\\
\hline
36\,h & \nsig{$0$} & $+30$ & $+30$ & \nsig{$+6$} & $+18$ & \nsig{$+6$} & \nsig{$0$} & $+18$ & \nsig{$+12$}\\
48\,h & \nsig{$0$} & $+36$ & $+24$ & \nsig{$+6$} & $+12$ & \nsig{$+6$} & \nsig{$0$} & $+24$ & \nsig{$+6$}\\
\hline
60\,h & \nsig{$+6$} & -- & $+24$ & \nsig{$0$} & -- & \nsig{$+6$} & \nsig{$0$} & -- & \nsig{$+6$}\\
72\,h & \nsig{$-6$} & -- & $+18$ & \nsig{$+6$} & -- & \nsig{$+6$} & \nsig{$-6$} & -- & \nsig{$+6$}\\
\hline
84\,h & \nsig{$-12$} & -- & $+18$ & \nsig{$+6$} & -- & \nsig{$0$} & \nsig{$-6$} & -- & \nsig{$+6$}\\
96\,h & \nsig{$-6$} & -- & $+12$ & \nsig{$0$} & -- & \nsig{$-6$} & \nsig{$-6$} & -- & \nsig{$0$}\\
\hline
108\,h & \nsig{$0$} & -- & \nsig{$+6$} & \nsig{$+6$} & -- & \nsig{$+6$} & \nsig{$-6$} & -- & \nsig{$+12$}\\
120\,h & \nsig{$+6$} & -- & \nsig{$+12$} & \nsig{$+12$} & -- & \nsig{$-6$} & \nsig{$+6$} & -- & \nsig{$+6$}\\
\hline
132\,h & \nsig{$0$} & -- & \nsig{$+12$} & \nsig{$+6$} & -- & \nsig{$+6$} & \nsig{$0$} & -- & \nsig{$0$}\\
144\,h & \nsig{$0$} & -- & \nsig{$+6$} & \nsig{$+6$} & -- & \nsig{$-6$} & \nsig{$-6$} & -- & \nsig{$+6$}\\
\hline
156\,h & \nsig{$0$} & -- & \nsig{$+6$} & \nsig{$+6$} & -- & \nsig{$+12$} & \nsig{$+6$} & -- & \nsig{$+6$}\\
168\,h & \nsig{$+12$} & -- & \nsig{$+6$} & \nsig{$0$} & -- & \nsig{$+12$} & \nsig{$0$} & -- & \nsig{$+12$}\\
\hline
180\,h & \nsig{$+6$} & -- & \nsig{$+18$} & \nsig{$+6$} & -- & \nsig{$0$} & \nsig{$0$} & -- & \nsig{$0$}\\
192\,h & \nsig{$+6$} & -- & \nsig{$+6$} & \nsig{$0$} & -- & \nsig{$-12$} & \nsig{$-12$} & -- & \nsig{$+18$}\\
\hline
204\,h & \nsig{$0$} & -- & \nsig{$-18$} & \nsig{$-12$} & -- & \nsig{$-12$} & \nsig{$+6$} & -- & \nsig{$-6$}\\
216\,h & \nsig{$-6$} & -- & \nsig{$-12$} & \nsig{$-18$} & -- & \nsig{$-24$} & \nsig{$-6$} & -- & \nsig{$-6$}\\
\hline
228\,h & \nsig{$+12$} & -- & \nsig{$+6$} & \nsig{$-30$} & -- & \nsig{$-6$} & \nsig{$+6$} & -- & \nsig{$+12$}\\
240\,h & \nsig{$+6$} & -- & \nsig{$+6$} & \nsig{$0$} & -- & \nsig{$0$} & \nsig{$0$} & -- & \nsig{$0$}\\
\hline
\end{tabular}
\caption{
    \textbf{Lead time gain for the atmospheric fields.}
    Otherwise, same as \cref{sitab:equivalent-surface}.
}
\label{sitab:equivalent-upper-air}
\end{table}

\clearpage
\subsection{Additional Regional Evaluation}
\label{si:extended-regional}

Here we show an extended evaluation of Nested-EAGLE, ML-GFS-Base (the
GFS-only baseline model, see \cref{subsec:baseline-design}), and GFS (Global
Forecast System).
We show RMSE for many variables and vertical levels for a number of different
regions across the globe.
As in the main text, we compute RMSE using in situ observations as a reference,
from 293~forecasts initialized every 30~hours during the test period:
February 2024--January 2025.
In all figures in this section, the solid lines indicate the median, and
shading indicates the 95\% confidence interval.
To simplify the presentation, we evaluate all models at 0.25$^\circ$
resolution, even for the evaluation over the
Contiguous United States (CONUS; \cref{sifig:conus-rmse}).

\begin{figure}[H]
    \centering
    \includegraphics[width=\textwidth]{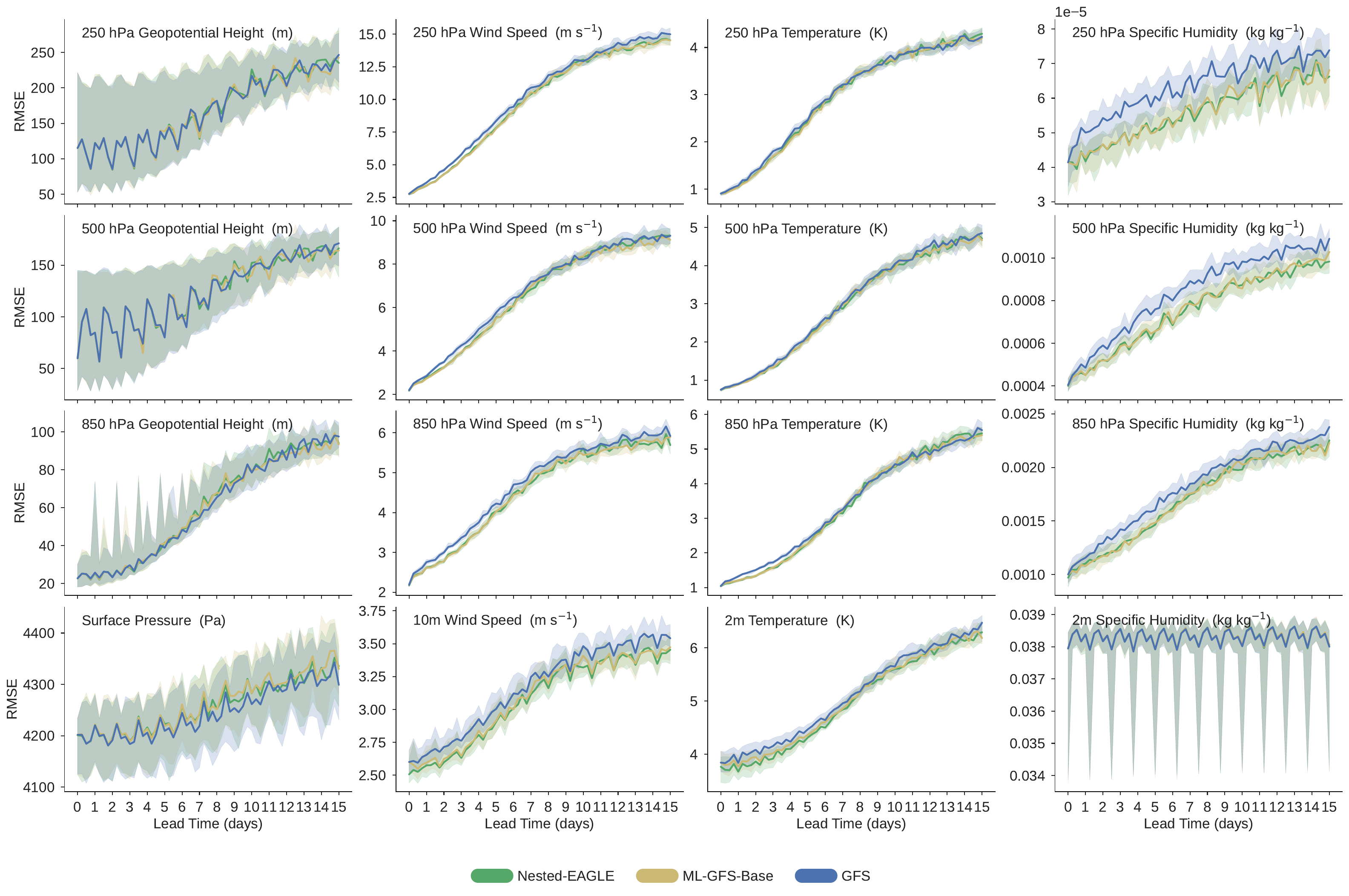}
    \caption{
        \textbf{
            Global RMSE.
        }
        RMSE against in situ observations over the globe.
    }
    \label{sifig:global-rmse}
\end{figure}

\begin{figure}[H]
    \centering
    \includegraphics[width=\textwidth]{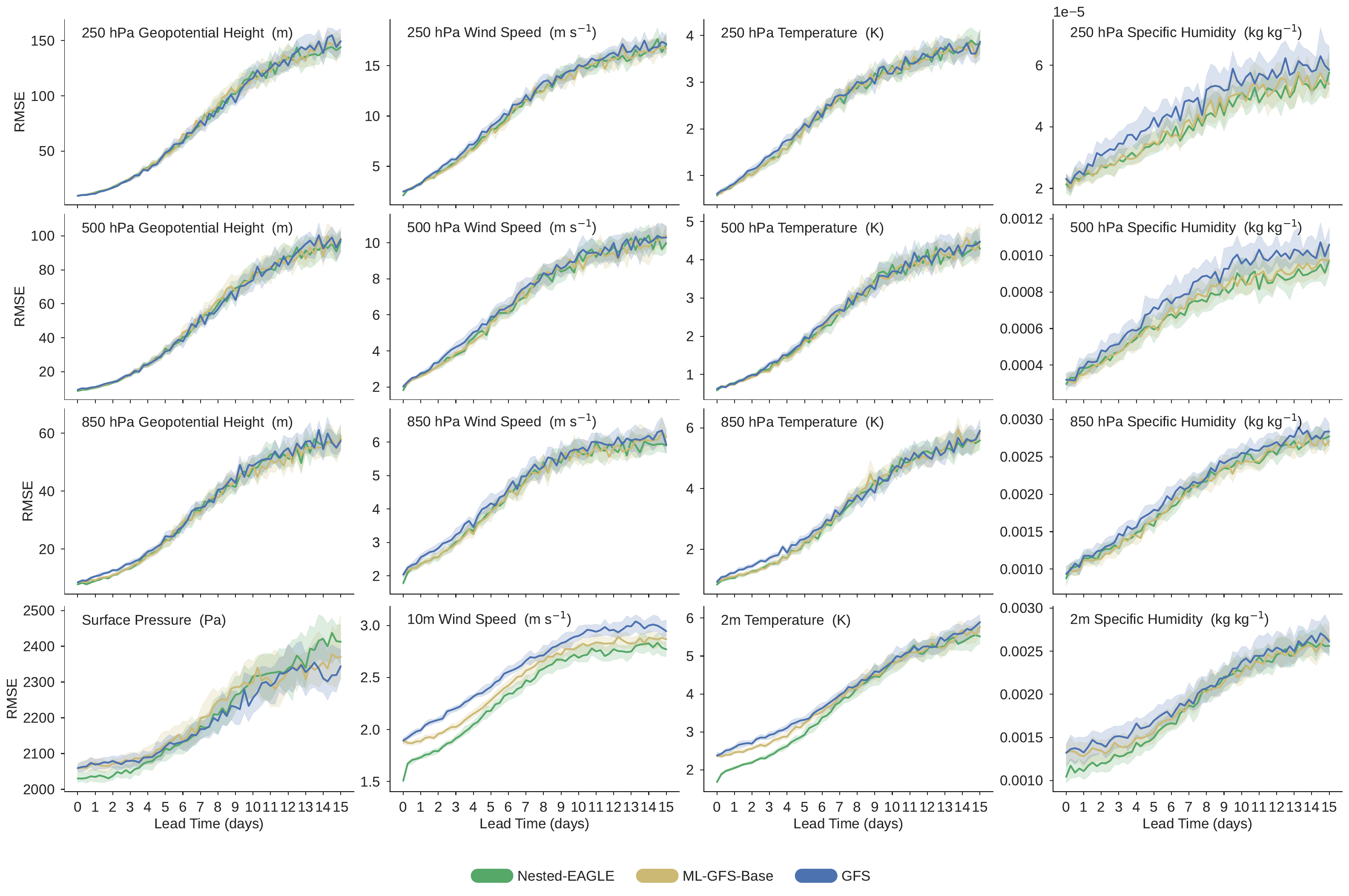}
    \caption{
        \textbf{
            CONUS RMSE.
        }
        Same as \cref{sifig:global-rmse}, except restricted to
        20$^\circ$N--55$^\circ$N and 135$^\circ$W--50$^\circ$W.
        The difference between this figure and \cref{fig:lam_rmse} is
        that more variables are shown here, and that this evaluation is carried
        out at 0.25$^\circ$ resolution, not 6~km.
    }
    \label{sifig:conus-rmse}
\end{figure}

\begin{figure}[H]
    \centering
    \includegraphics[width=\textwidth]{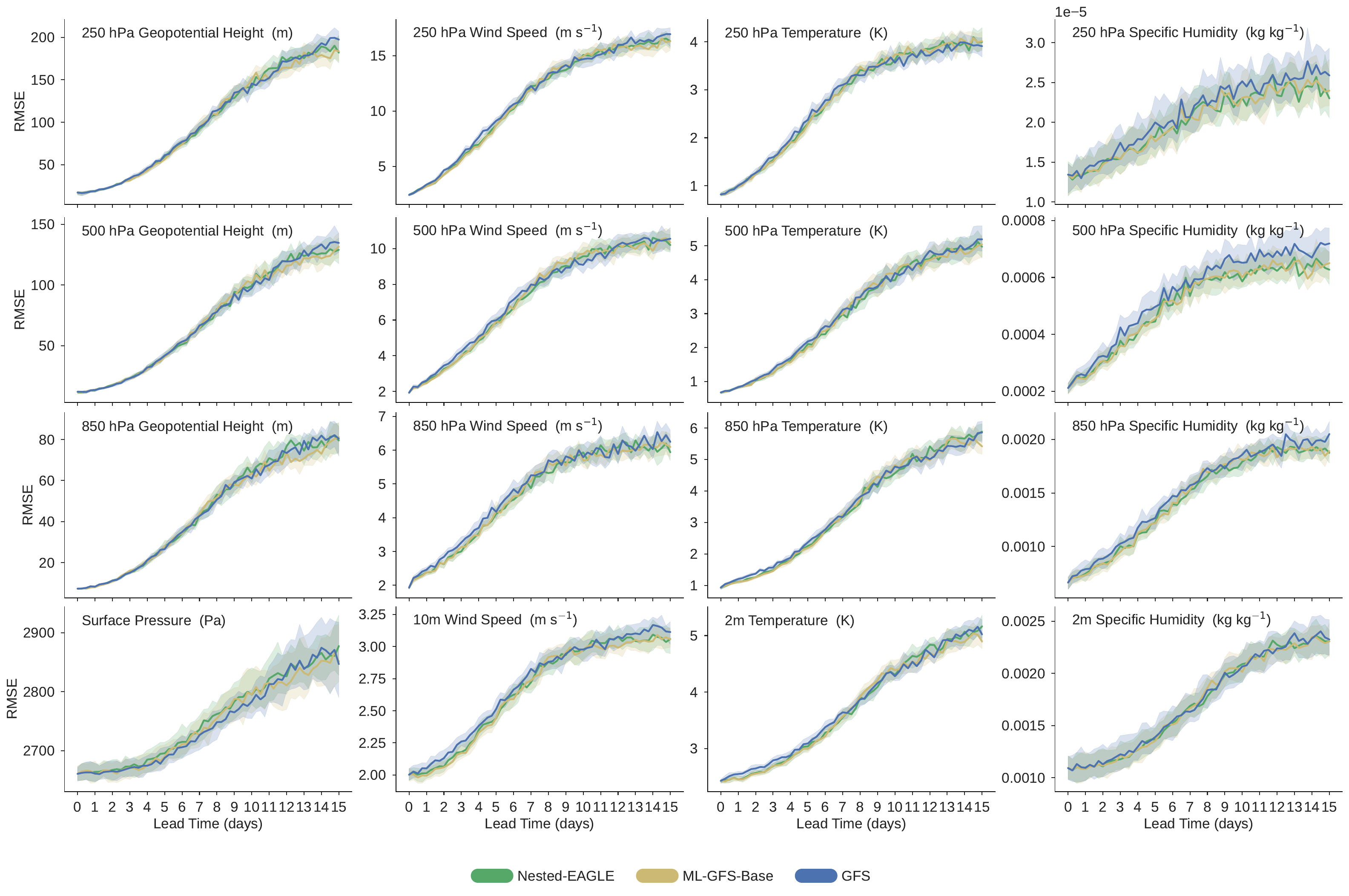}
    \caption{
        \textbf{
            Europe RMSE.
        }
        Same as \cref{sifig:global-rmse}, except restricted to
        35$^\circ$N--75$^\circ$N and 25$^\circ$W--65$^\circ$E.
        The only difference between this figure and \cref{fig:europe_rmse} is
        that more variables are shown here.
    }
    \label{sifig:europe-rmse}
\end{figure}

\begin{figure}[H]
    \centering
    \includegraphics[width=\textwidth]{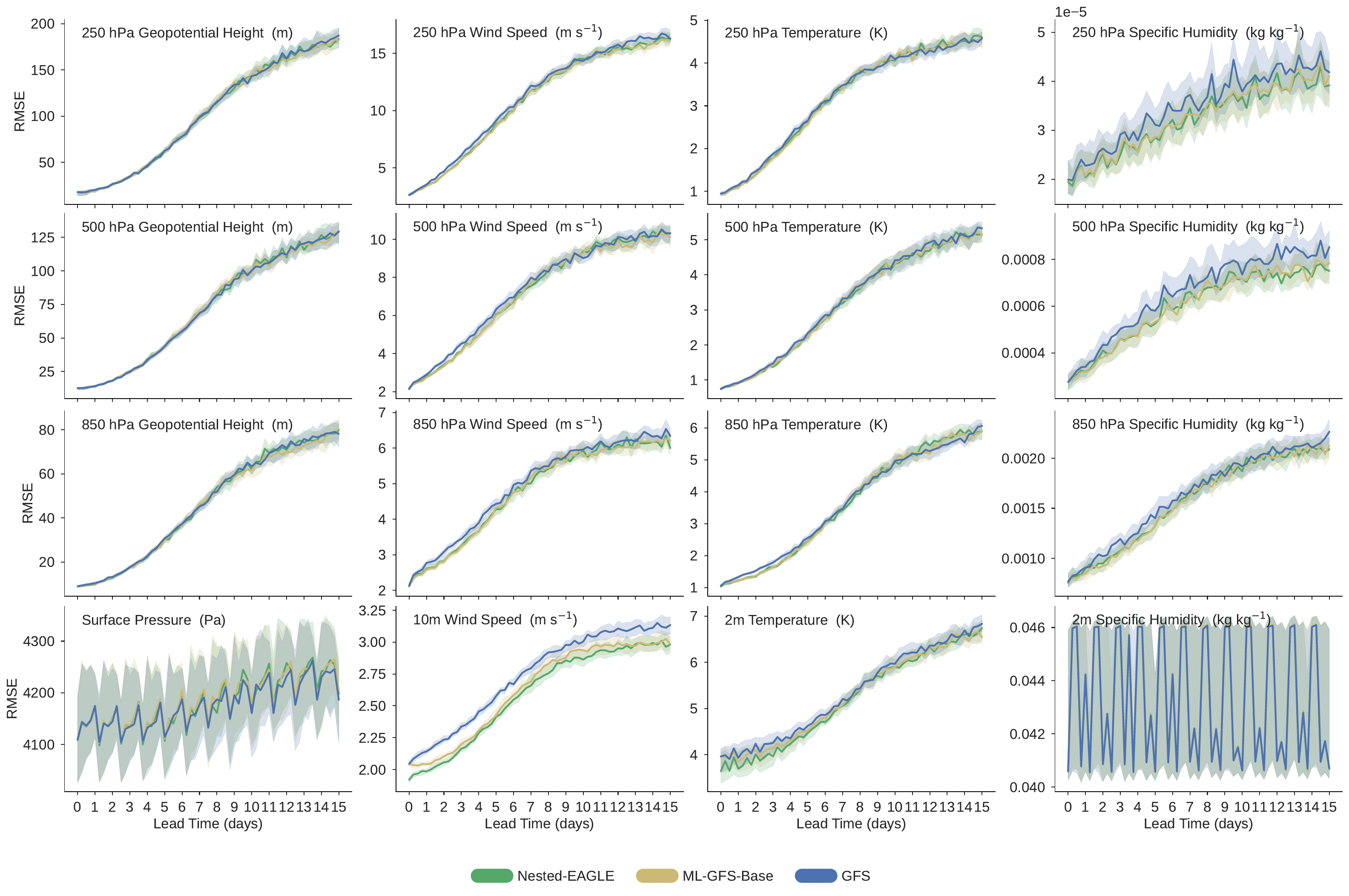}
    \caption{
        \textbf{
            Northern Hemisphere RMSE.
        }
        Same as \cref{sifig:global-rmse}, except restricted to
        20$^\circ$N--80$^\circ$N.
    }
    \label{sifig:nh-rmse}
\end{figure}

\begin{figure}[H]
    \centering
    \includegraphics[width=\textwidth]{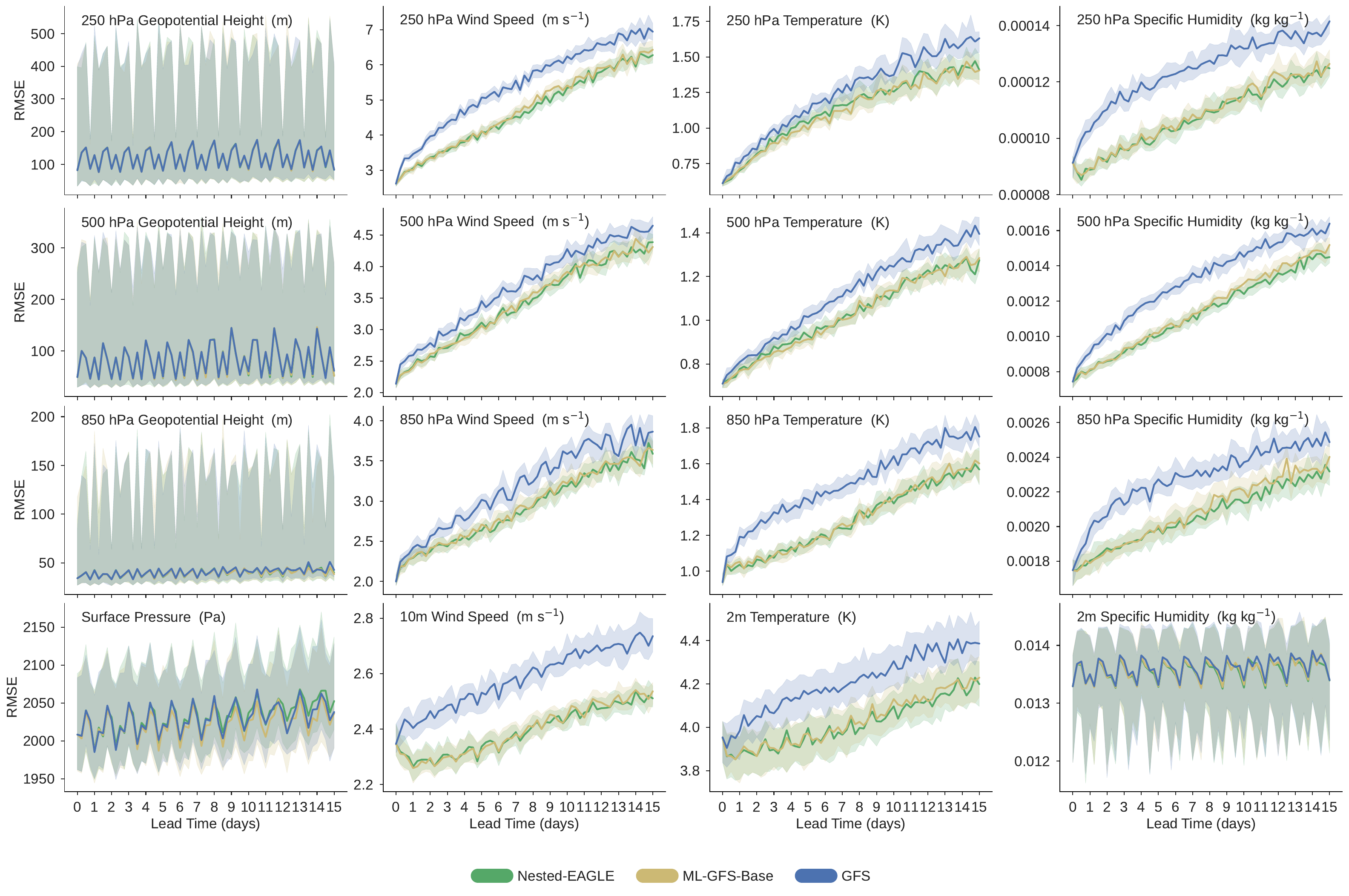}
    \caption{
        \textbf{
            Tropics RMSE.
        }
        Same as \cref{sifig:global-rmse}, except restricted to
        20$^\circ$S--20$^\circ$N.
    }
    \label{sifig:tropics-rmse}
\end{figure}

\begin{figure}[H]
    \centering
    \includegraphics[width=\textwidth]{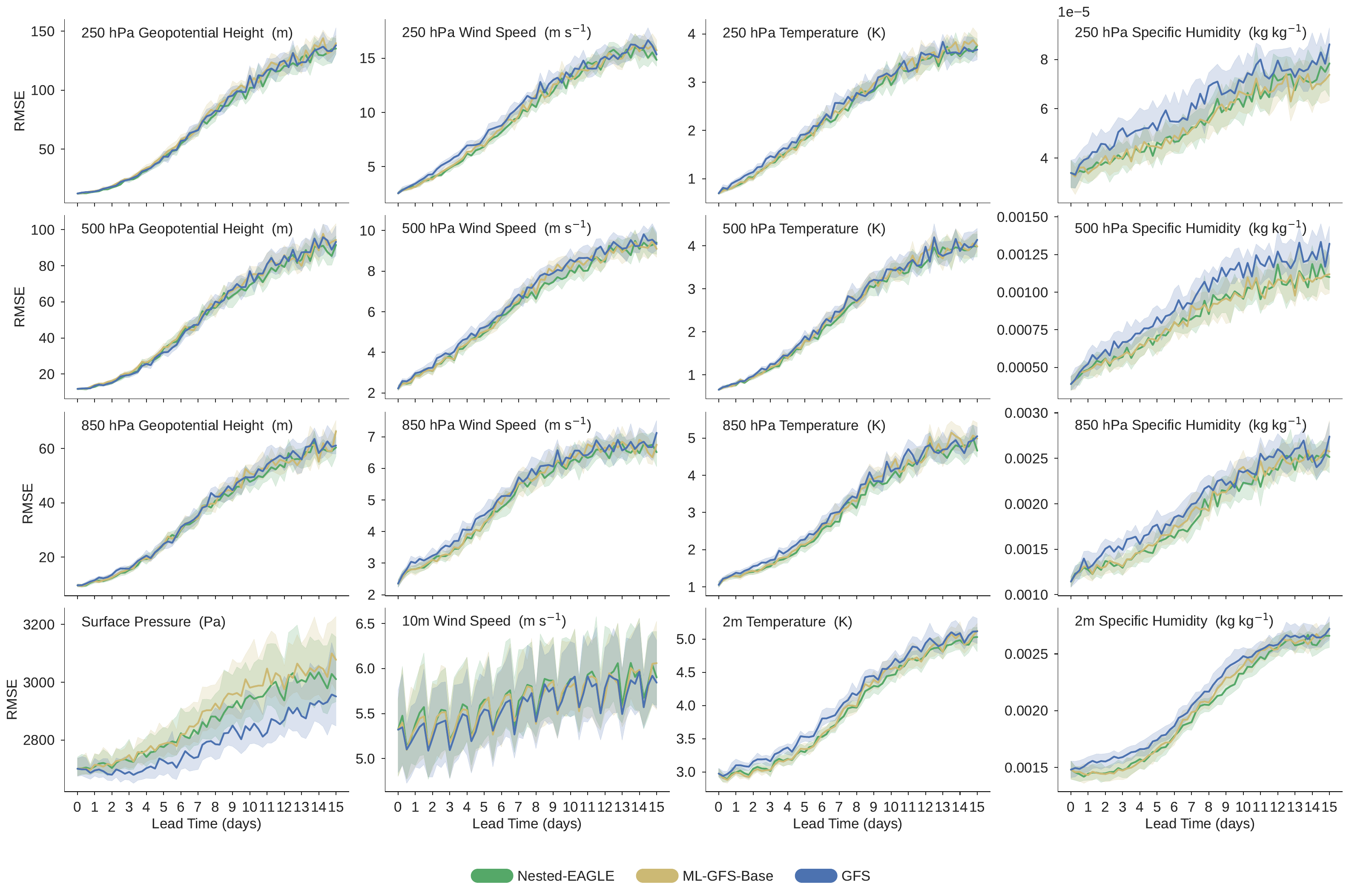}
    \caption{
        \textbf{
            Southern Hemisphere RMSE.
        }
        Same as \cref{sifig:global-rmse}, except restricted to
        80$^\circ$S--20$^\circ$S.
    }
    \label{sifig:sh-rmse}
\end{figure}

\begin{figure}[H]
    \centering
    \includegraphics[width=\textwidth]{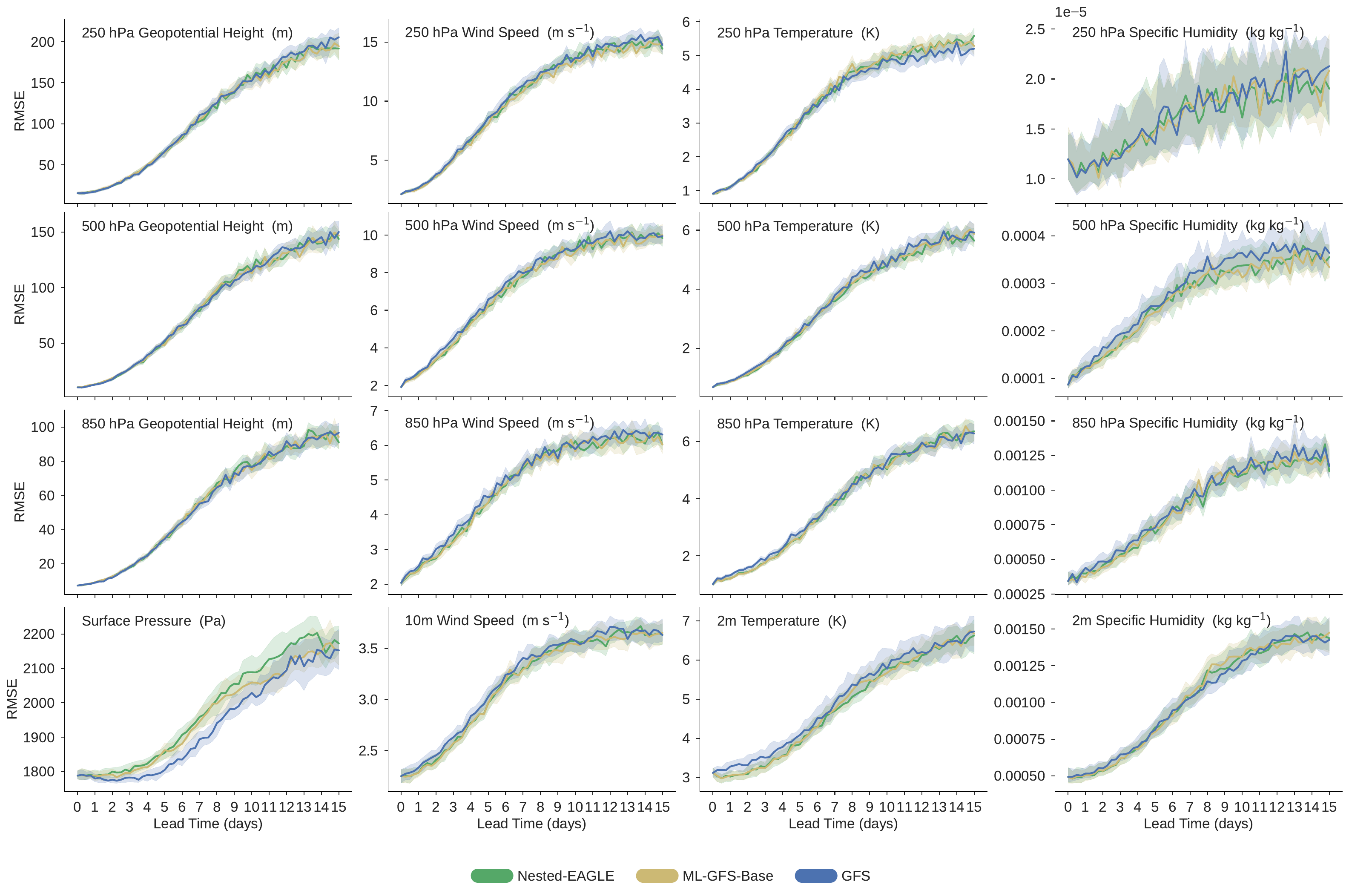}
    \caption{
        \textbf{
            Polar North RMSE.
        }
        Same as \cref{sifig:global-rmse}, except restricted to
        60$^\circ$N--90$^\circ$N.
    }
    \label{sifig:polar-north-rmse}
\end{figure}

\begin{figure}[H]
    \centering
    \includegraphics[width=\textwidth]{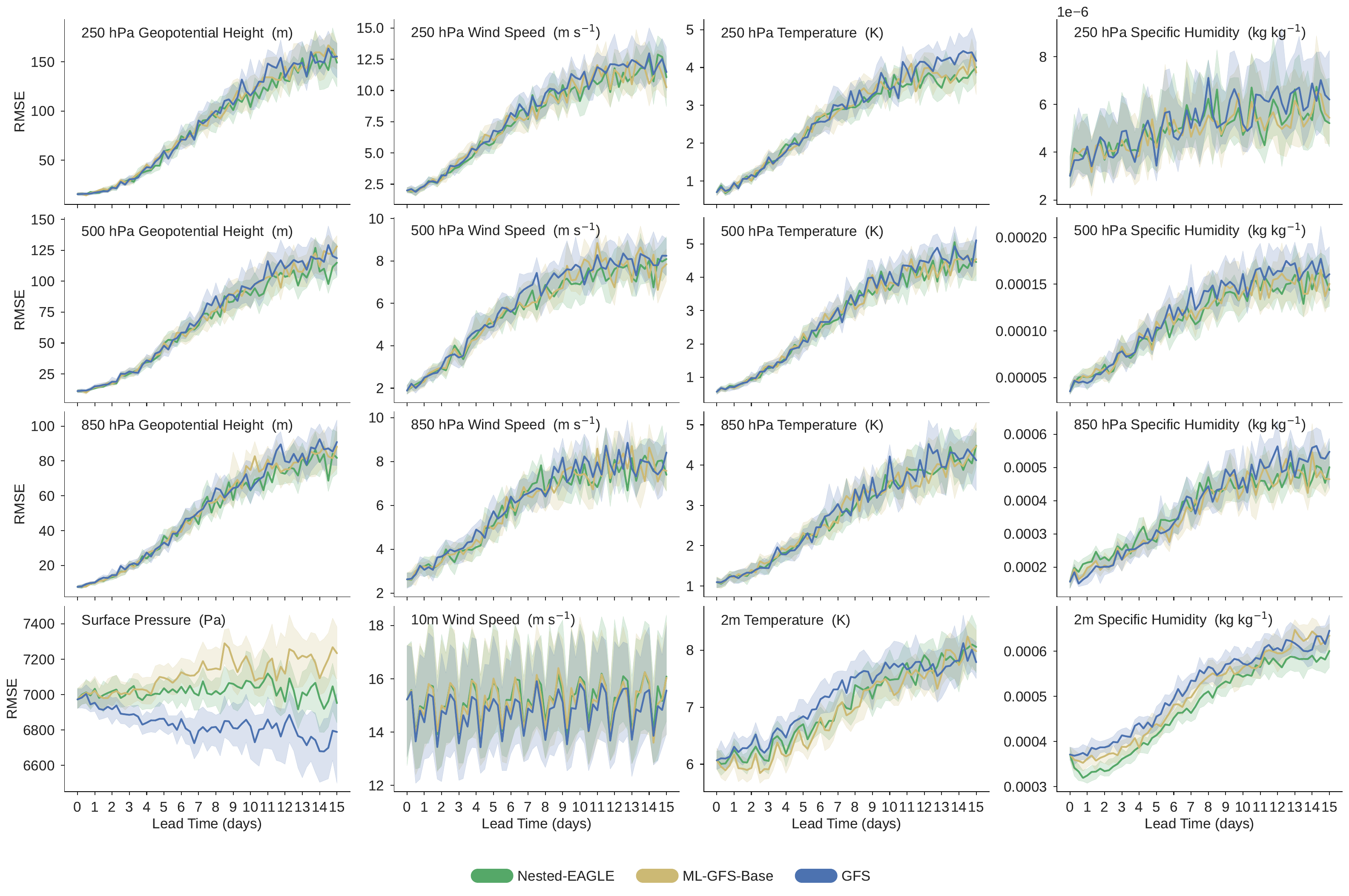}
    \caption{
        \textbf{
            Polar South RMSE.
        }
        Same as \cref{sifig:global-rmse}, except restricted to
        90$^\circ$S--60$^\circ$S.
    }
    \label{sifig:polar-south-rmse}
\end{figure}

\clearpage
\section{Model Development Details and Ablation Experiments}
\label{si:model-dev}

Here we present experimental results that drove development decisions for
Nested-EAGLE.
For development purposes, and in all of the results shown in this section, we
used coarser data than those shown in the main text.
Specifically, we conservatively regridded High-Resolution Rapid Refresh (HRRR)
data to 15~km and GFS data to
1$^\circ$ resolution.
Aside from \cref{si:training-steps}, we show results from models
trained for 30{,}000 optimization iterations, using only a single
6~hour forecast step in the loss function
(i.e., corresponding to Stage A in \cref{table:training}).
At this coarser resolution and with the sliding-window transformer processor described in
\cref{subsec:nested-eagle-design}, this development model was relatively cheap
to train, requiring only $\sim$4.5~hours of wall-clock time across 8~GPU nodes on
Perlmutter.
This inexpensive model enabled many tests, some of which are outlined here.
We make the necessary assumption that the results shown here
carry over to the higher-resolution model.

All evaluation results shown in this section are based on 158 model forecasts
initialized every 54~hours during the
validation period: February 2023--January 2024.
While the results in the main text show evaluation against in situ observations,
some of the results here are based on errors against HRRR analysis, since the
observation evaluation pipeline was not in place when all experiments were
carried out.
Additionally, the models shown here sometimes have different
hyperparameters than the final Nested-EAGLE design, but, importantly, the only
hyperparameter that changes in each figure is the one highlighted in the plot.
Once again, we make the necessary assumption that these results show general trends that
carry over, even with other changes to the model.

\subsection{LAM Loss Fraction}
\label{si:lam-loss}

\begin{figure}[H]
    \centering
    \includegraphics[width=\textwidth]{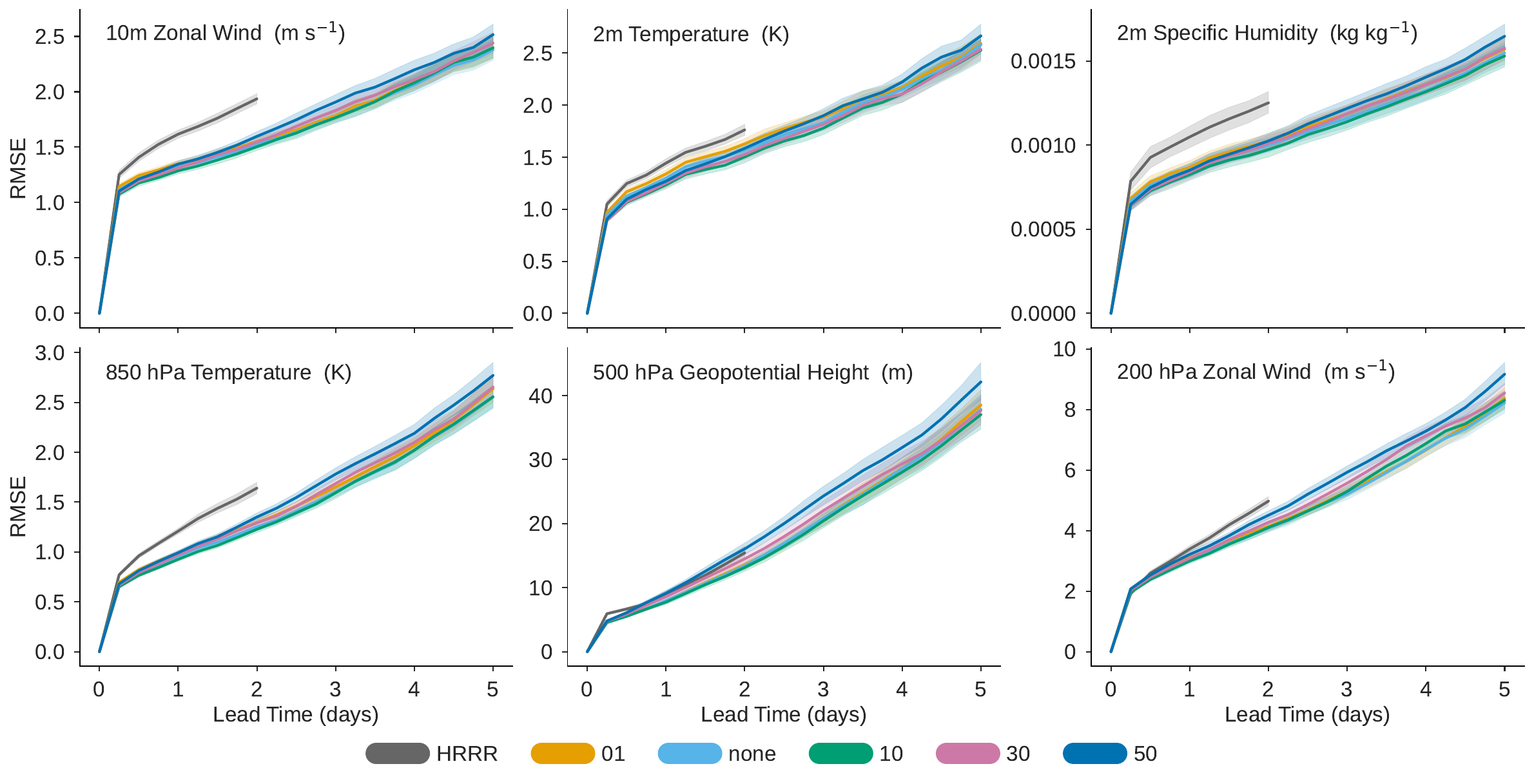}
    \caption{
        \textbf{
            LAM loss fraction sensitivity tests.
        }
        RMSE against HRRR analysis.
        Each color indicates the percentage of loss dedicated to the LAM region
        (i.e., HRRR subdomain) during training.
        Note that ``none'' refers to the case where no reweighting is
        performed, so the region takes on a weighting based purely on the
        spatial extent of the HRRR domain relative to the globe, which is about 3.4\%.
        \meanci{158}
    }
    \label{sifig:lam-loss}
\end{figure}

A key hyperparameter in nested or stretched grid models is the priority given to
the target Limited Area Model (LAM) region during training.
Following \citet{nipen_regional_2026} we set this priority by reweighting the
loss function, so that the target region attains a certain percentage of the
total loss, allowing it to have more influence than it otherwise would with the
grid-cell area weighting.
\cref{sifig:lam-loss} shows the skill of models with different settings for this
hyperparameter.
Clearly, 50\% is too high, but RMSE values for ``none'', 10\%, and 30\% are actually
quite close.
We selected 10\% as a final value, as it tended to produce the best
results.

\subsection{Attention Window Size}
\label{si:window-size}

\begin{figure}[H]
    \centering
    \includegraphics[width=\textwidth]{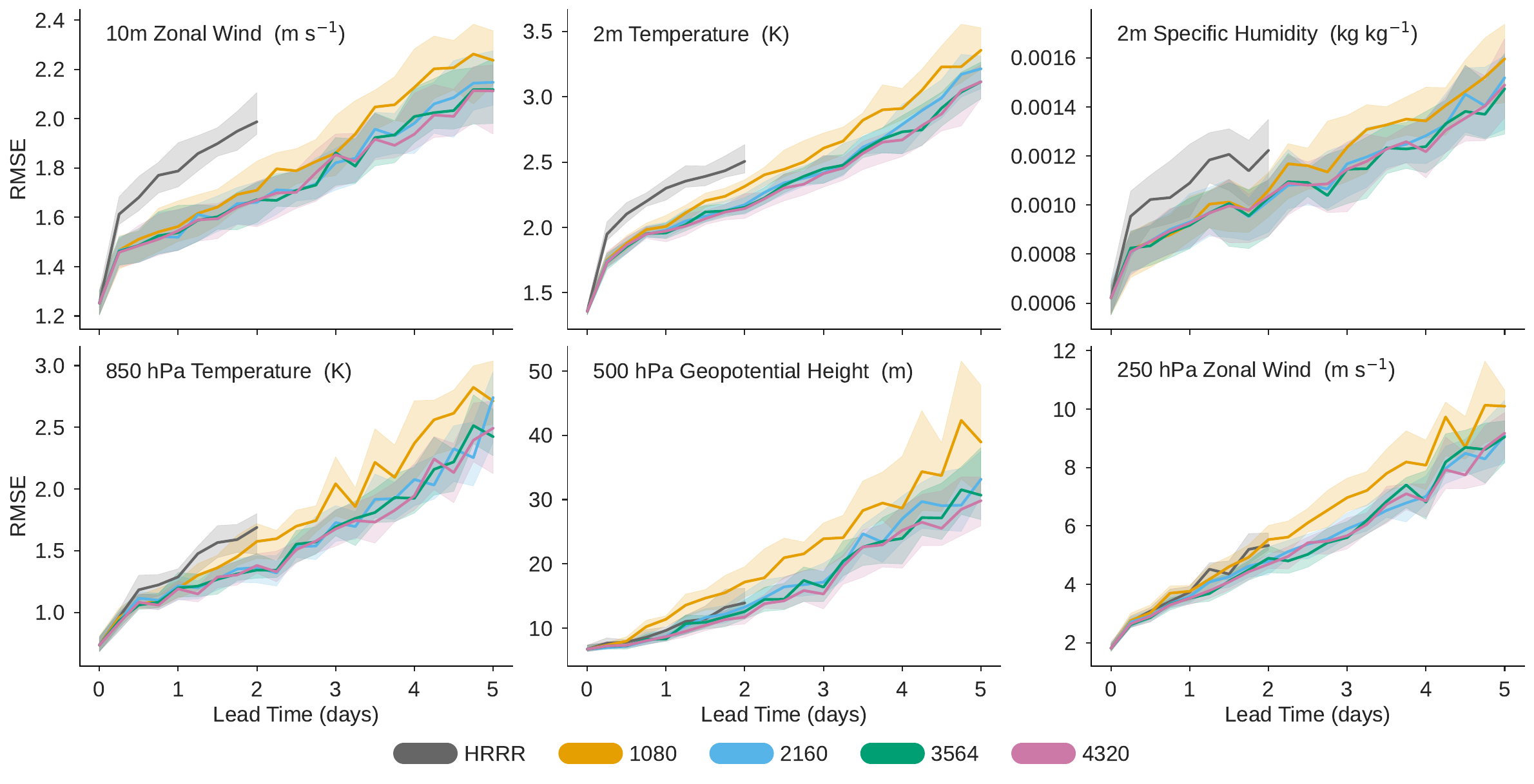}
    \caption{
        \textbf{
            Attention window size sensitivity test.
        }
        RMSE against in situ observations.
        The colors indicate the number of latent mesh nodes in the westward and
        eastward directions from a query point during attention computations
        (see visualization in \cref{fig:nested_illustration}).
        \medianci{158}
    }
    \label{sifig:window-size}
\end{figure}

An essential hyperparameter for the sliding-window transformer architecture is the
attention window size, which determines the number of neighboring points
included during attention computations for a given query point.
With a multi-resolution model like Nested-EAGLE, it is not immediately clear how
the window size maps to physical distances, since grid nodes have different spacings for
different regions of the globe.
See \cref{fig:nested_illustration} for a visualization, where latent nodes
during a single attention computation are highlighted in blue.
Thus, we rely on empirical results to drive our decision.
\cref{sifig:window-size} shows the forecast skill over CONUS as a function of window size.
Recall that the numbers presented in \cref{sifig:window-size} correspond to a much smaller
latent space than for the higher-resolution Nested-EAGLE.

Clearly, a window size of 1{,}080 is too small, but differences for higher
values are largely statistically insignificant.
We opted for 3{,}564 based on an ad hoc rule of thumb to guide the
process of scaling up the design to higher resolution.
That is, 3{,}564 is the number of latent mesh nodes, divided by twice the
number of processor stages:
$N_\text{window} \coloneqq N_\text{mesh} / N_\text{proc} / 2$.
Given this rule, our final model implementation used a window size of 8{,}168.
Also influencing this decision was the fact that we
observed little computational sensitivity to the window size parameter during
development.
We presume that the lack of sensitivity is due
to the overwhelming time spent loading data into memory, which requires
reading two datasets for the nested setup, and to the extremely high efficiency of FlashAttention
\citep{dao_flashattention_2022,dao_flashattention-2_2023}.

\subsection{Training Iterations}
\label{si:training-steps}

\begin{figure}[H]
    \centering
    \includegraphics[width=\textwidth]{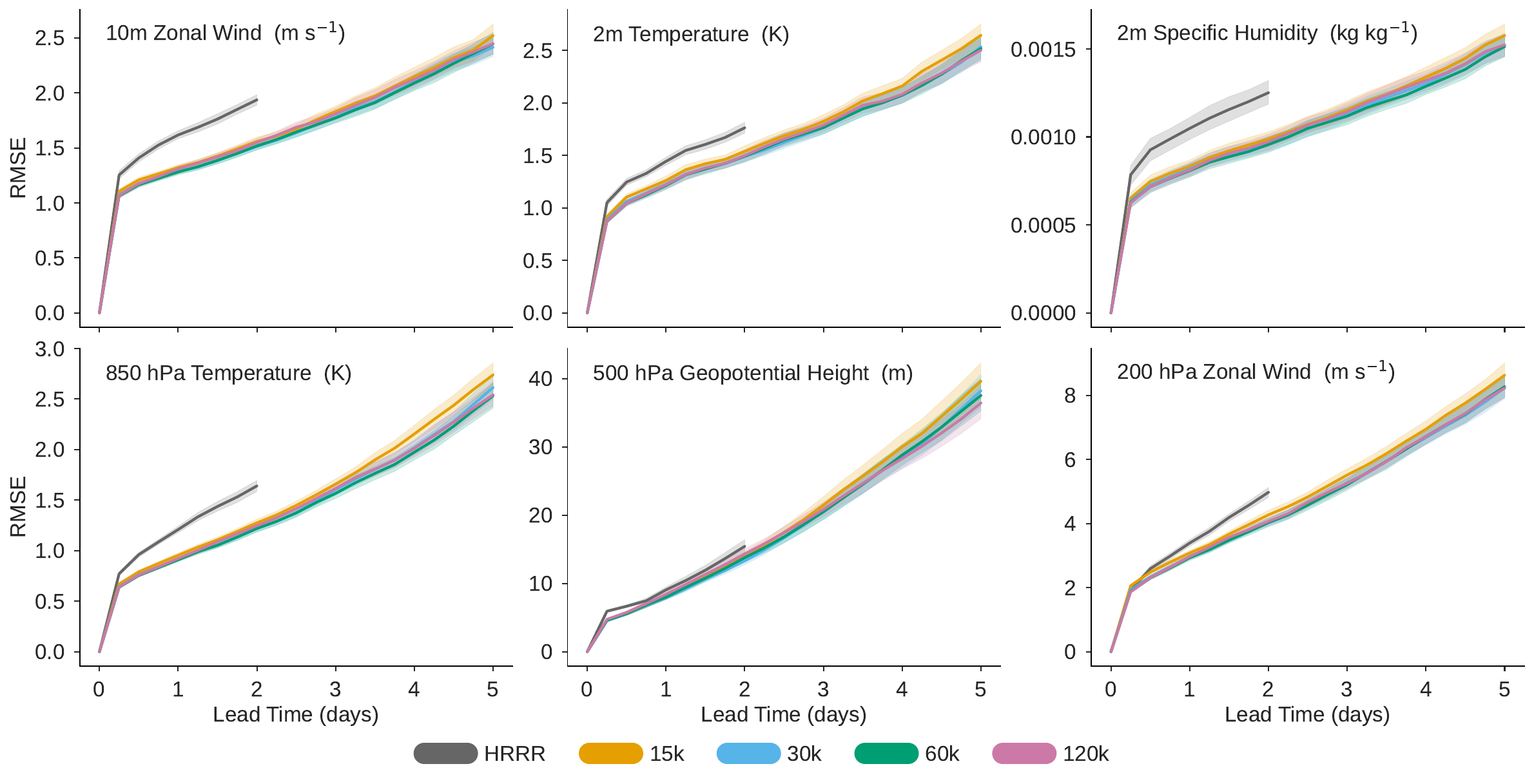}
    \caption{
        \textbf{
            Training iterations sensitivity test.
        }
        RMSE against HRRR analysis.
        The colors indicate the number of optimization iterations taken during
        training, where each iteration consists of
        a single 6~hour forecast step.
        \meanci{158}
    }
    \label{sifig:training-steps}
\end{figure}

\cref{sifig:training-steps} shows how the number of training iterations influences
forecast skill, only focusing on the single-step training phase (i.e., Stage A
in \cref{table:training}).
Note that each experiment indicates a distinct training cycle, completing
the learning rate schedule in the number of iterations indicated.
While 60{,}000 iterations produce the best results, the gains are
insignificant compared to models trained for 30{,}000 iterations,
although 15{,}000 iterations appear to be too short.
These results motivated using a schedule of 30{,}000 training iterations during
Stage A for all development experiments, and 60{,}000 training iterations during
Stage A for the final model.

We note that our model requires only about 10--20\% of the training iterations
that others use
\citep[e.g.,][]{lam_learning_2023,lang_aifs_2024}.
We presume that this is at least in part due to the fact that our training dataset is
only 8~years long, whereas the ERA5 dataset \citep{hersbach_era5_2020} used for a similar level of training
by \citet{lam_learning_2023} is $\sim$40~years long, about 5 times the size.

\subsection{Number of Latent Space Channels}
\label{si:latent-width}

\begin{figure}[H]
    \centering
    \includegraphics[width=\textwidth]{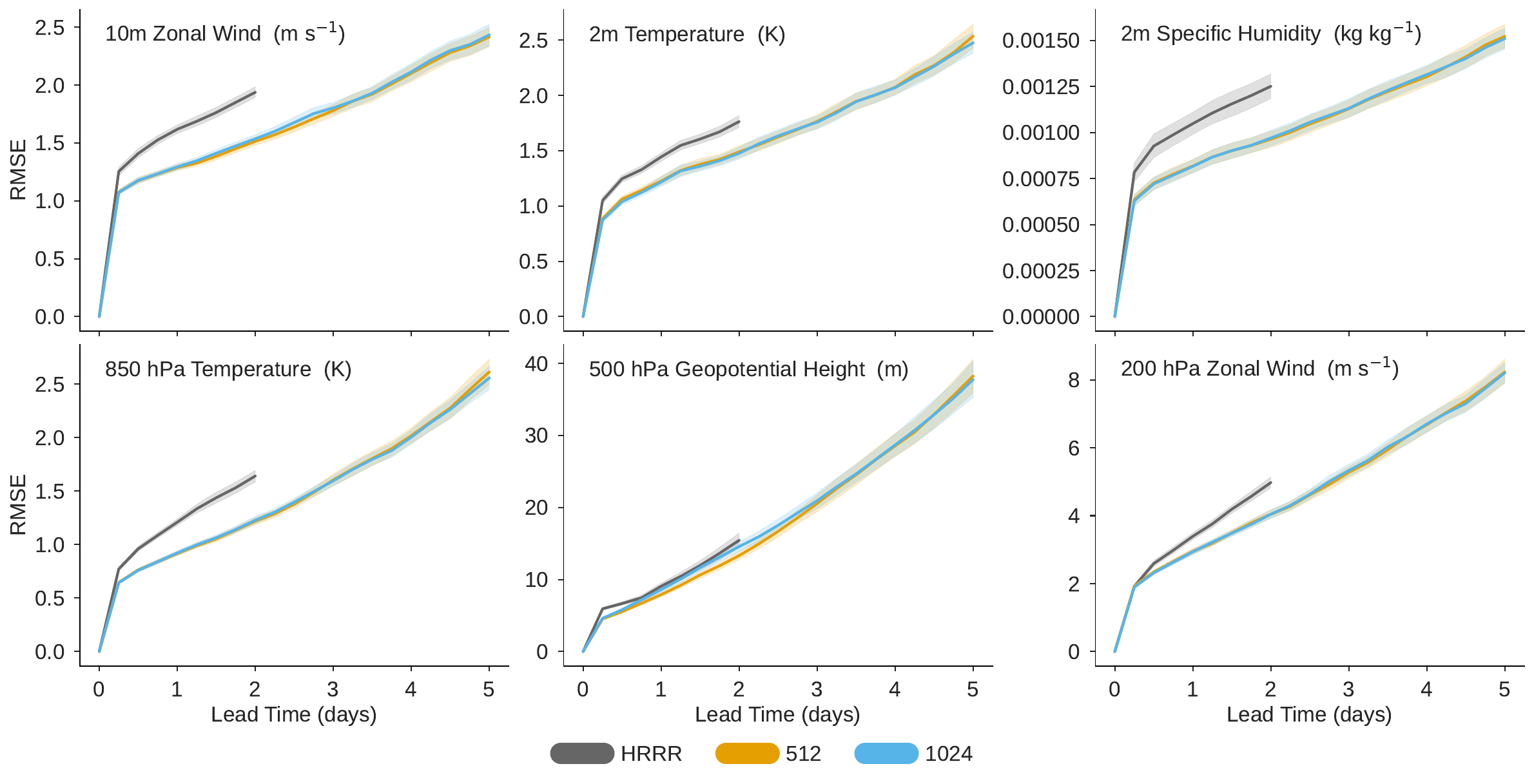}
    \caption{
        \textbf{
            Latent channels sensitivity test.
        }
        RMSE against HRRR analysis.
        The colors indicate the number of channels used in the latent space,
        i.e., the width of the neural network.
        \meanci{158}
    }
    \label{sifig:latent-width}
\end{figure}

Nested-EAGLE uses a relatively small latent space compared to Bris \citep{nipen_regional_2026}
and the Artificial Intelligence Forecasting System \citep[AIFS;][]{lang_aifs_2024}, all developed
in \texttt{anemoi}.
\cref{sifig:latent-width} shows the justification for this choice: we found no
statistical difference between 512 and 1{,}024 latent channels (i.e., the
width of the latent space vector), and yet the computational cost scales
linearly with this parameter.
We surmise that we see little benefit because our 8~year training dataset size
is relatively small compared to those used by other machine learning weather prediction
models, and so there are no effective
gains from the larger network.
We note that there was no benefit to training longer with the larger network,
although we do not show this explicitly.

\subsection{(No) Pressure Scaling in the Loss Function}
\label{si:pressure-scaling}

\begin{figure}[H]
    \centering
    \includegraphics[width=\textwidth]{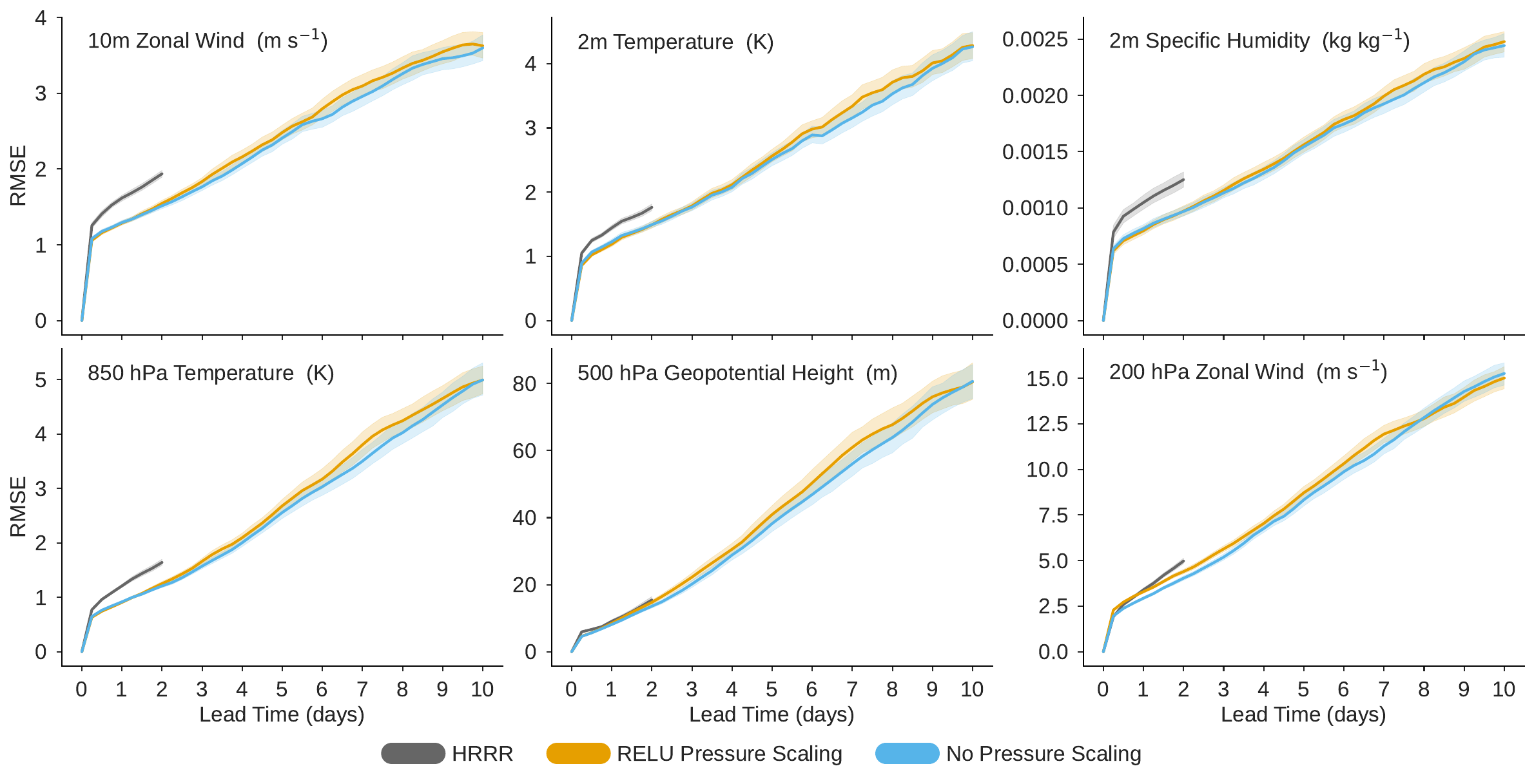}
    \caption{
        \textbf{
            Pressure scaling sensitivity test.
        }
        RMSE against HRRR analysis.
        The colors indicate whether pressure scaling was used in the loss
        function or not.
        \meanci{158}
    }
    \label{sifig:pressure-scaling}
\end{figure}

Unlike models like GraphCast and AIFS
\citep{lam_learning_2023,lang_aifs_2024},
Nested-EAGLE does not use a linear scaling with pressure in the loss function.
That is, we treat all vertical (pressure) levels identically in the loss
function, and do not prioritize lower levels over the upper levels.
\cref{sifig:pressure-scaling} shows our justification for this choice.
Not only do we see better performance for variables in the middle and upper
troposphere, as one would expect, but we see that removing the pressure scaling also has
no impact or even \textit{improves} skill in low-level temperature and near-surface
variables.
While we do not have a clear understanding of why we see this benefit at lower levels, at
the very least these results motivate other developers to test this choice for their
application.

\subsection{Number of Initial States}
\label{si:nic}

\begin{figure}[H]
    \centering
    \includegraphics[width=\textwidth]{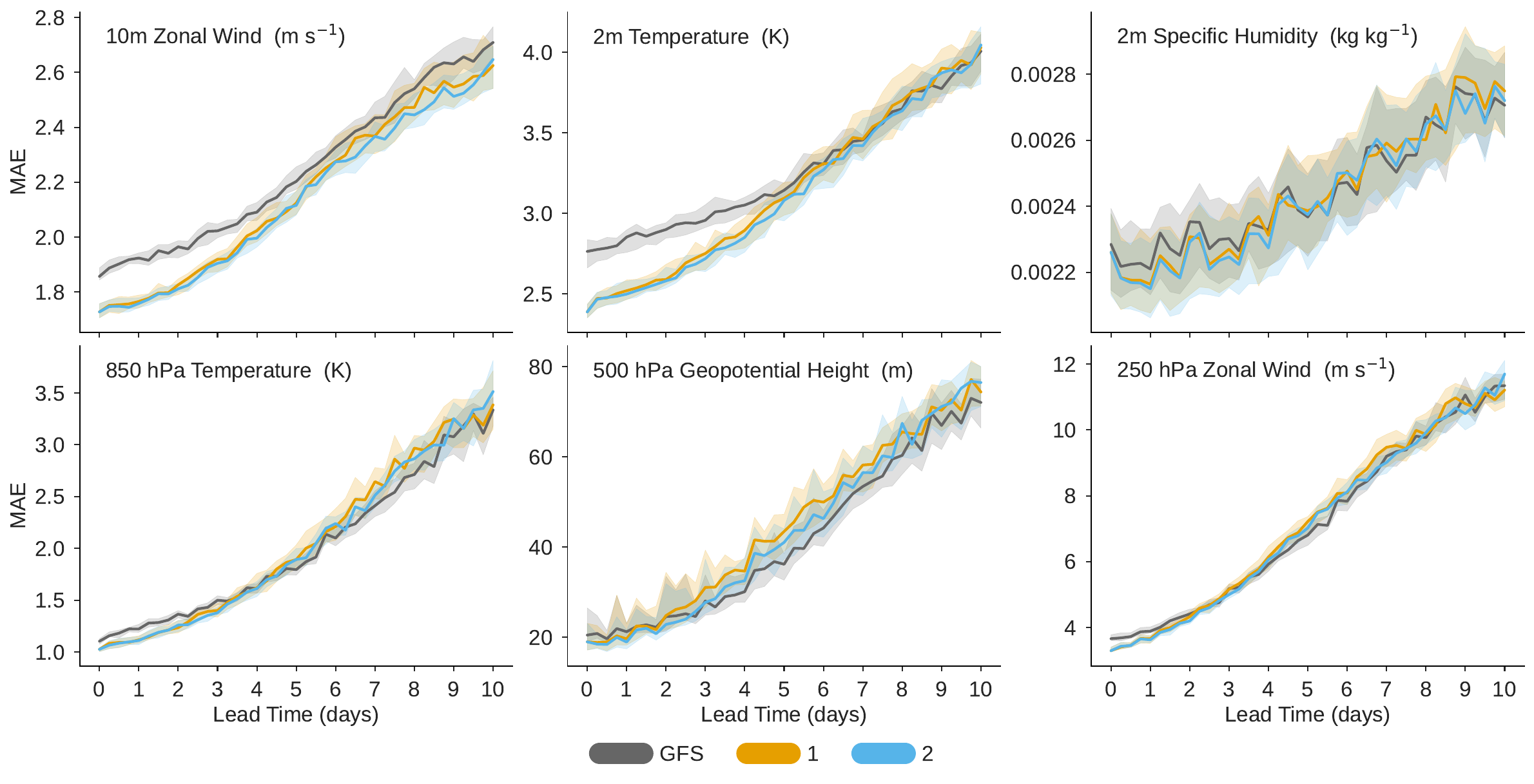}
    \caption{
        \textbf{
            Sensitivity to number of initial states.
        }
        MAE against in situ observations.
        The color indicates the number of initial states used.
        We use MAE here since it is less sensitive to outliers that dominate the 2m specific
        humidity signal.
        \medianci{158}
    }
    \label{sifig:nic}
\end{figure}

Following many others
\citep[e.g.,][]{lam_learning_2023,lang_aifs_2024,nipen_regional_2026}
Nested-EAGLE uses two initial states
(i.e., $\state(t)$ and $\state(t-6$~hours$)$) to produce a forecast
(\cref{eq:model}).
However, during the model development we found less sensitivity to this
hyperparameter than we expected, and therefore report the results here for the
benefit of others.
Specifically, \cref{sifig:nic} shows the impact on global Mean Absolute Error
(MAE) against in situ
observations from one or two initial states.
While it appears that using two initial states consistently improves MAE across
different variables, the differences are usually not statistically significant.
We chose to use two initial states for the final model design essentially
due to the gains in 500~hPa geopotential height skill when using two initial
states.
We reason that this is an important field for a public-facing weather model,
and so the extra costs of using two initial states are warranted.
However, these results highlight that using one initial state is probably
acceptable for many applications, especially during the early stages of model development.

\clearpage
\section{Attribution of Perturbed 2m Temperature Bias to Orography Differences}
\label{si:t2m-dz-attribution}

\cref{subsec:ic_experiment} showed that Nested-EAGLE(GFS~IC) rapidly approaches
the skill of Nested-EAGLE, but errors converge slightly more slowly with 2m
temperature.
Here we show evidence which suggests that this slower convergence is due to
orographic differences between the two configurations.
Recall that Nested-EAGLE(GFS~IC) is Nested-EAGLE
running inference with only GFS initial conditions and GFS static variables,
such as orography, whereas Nested-EAGLE uses HRRR initial conditions and orography over CONUS.

\begin{figure}[H]
    \centering
    \includegraphics[width=\textwidth]{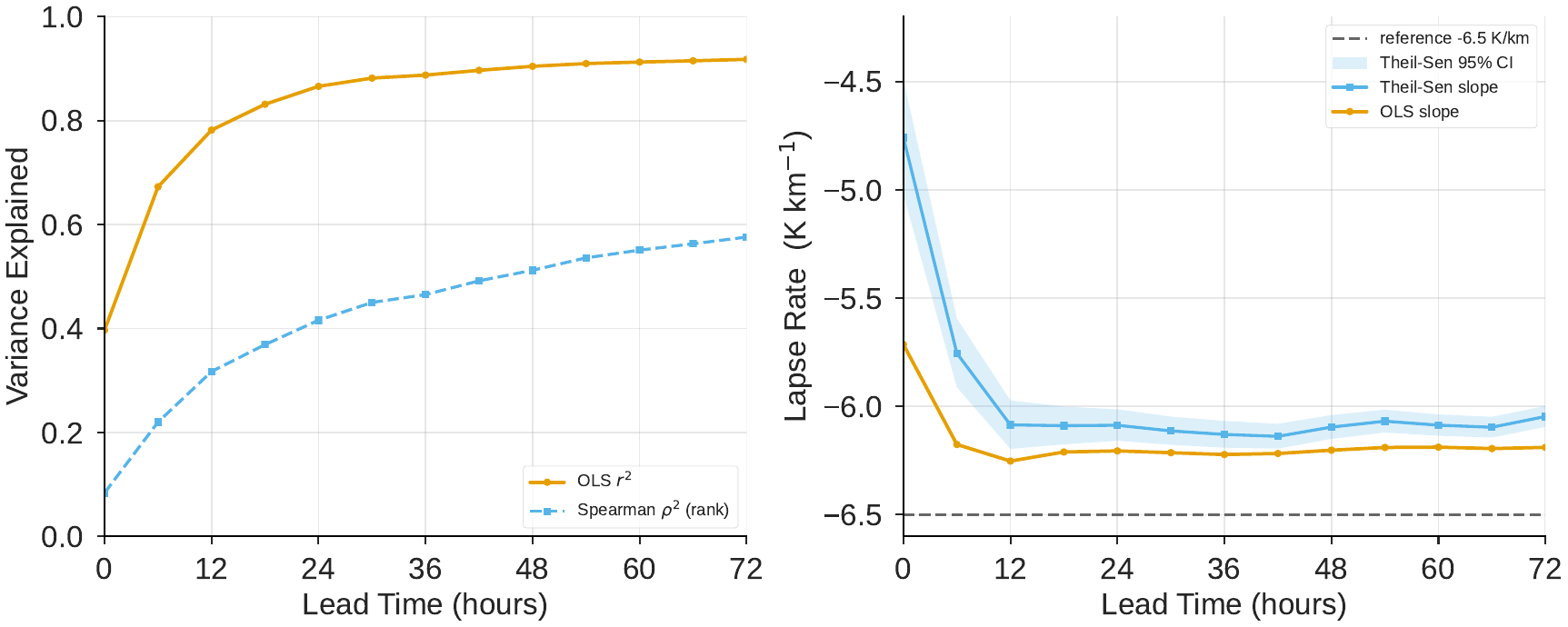}
    \caption{
        \textbf{
            Attribution of perturbed 2m temperature bias to orography
            differences.
        }
        The left panel shows Pearson's $r^2$ from an OLS regression and Spearman's
        $\rho^2$, where both indicate the correlation between 2m temperature
        bias (i.e., signed error averaged over initial conditions) and orography
        differences between Nested-EAGLE and Nested-EAGLE(GFS~IC).
        The right panel shows the lapse rates implied by the regression analyses,
        which are fairly close to the standard reference value in the troposphere.
        We computed the OLS regression over all 6~km grid cells, whereas we
        evaluated the Theil-Sen fit over a random sample of 8{,}000 grid cells,
        from which we estimated the 95\% confidence interval.
    }
    \label{sifig:t2m-bias-attribution}
\end{figure}

\cref{sifig:t2m-bias-attribution} shows that the average error, i.e., bias,
between Nested-EAGLE and Nested-EAGLE(GFS~IC) is highly correlated with
orography differences.
The left panel shows Pearson's $r^2$ from an Ordinary Least Squares (OLS) regression
between the temperature bias, averaged over forecast initializations, and the
signed difference between HRRR and GFS orography.
To be explicit, bias and orography differences as a function of location $i$ are defined as
\begin{equation}
    \text{Bias}(i,\Delta t) = \dfrac{1}{N_f}\sum_{j=1}^{N_f}
    \left(
        \pred_{\text{Nested-EAGLE}}(i, j, \Delta t) -
        \pred_{\text{Nested-EAGLE(GFS IC)}}(i,j,\Delta t)
    \right) \, ,
\end{equation}
\begin{equation}
    \Delta z(i) = z_\text{HRRR}(i) - z_\text{GFS}(i),
\end{equation}
where $j\in\{1, 2, ..., N_f\}$ denotes each forecast initialization out of
$N_f=293$.
As lead time increases, correlation increases rapidly over the first
day, and saturates at $r^2\simeq0.9$, indicating that the orography differences
are a leading cause of systematic errors.
Alongside $r^2$ we also show Spearman's $\rho^2$, which takes a more modest
value and rises more slowly, but still remains substantial, approaching $\rho^2\simeq0.6$ by 3~days.
We show both coefficients to indicate the importance of large
elevation differences in the correlation analysis, especially over the
mountainous western U.S.
The right panel of \cref{sifig:t2m-bias-attribution} shows an approximate lapse
rate implied by the regressions summarized in the left panel.
The OLS and Theil--Sen slopes level off at about $-6$~K~km$^{-1}$, which is
reasonably close to the standard tropospheric lapse rate of $-6.5$~K~km$^{-1}$.

\clearpage
\section{Extended Precipitation Evaluation}
\label{si:extended-precip}

\begin{figure}[H]
    \centering
    \includegraphics[
        width=\textwidth,
        trim=0 0 0 32,
        clip,
    ]{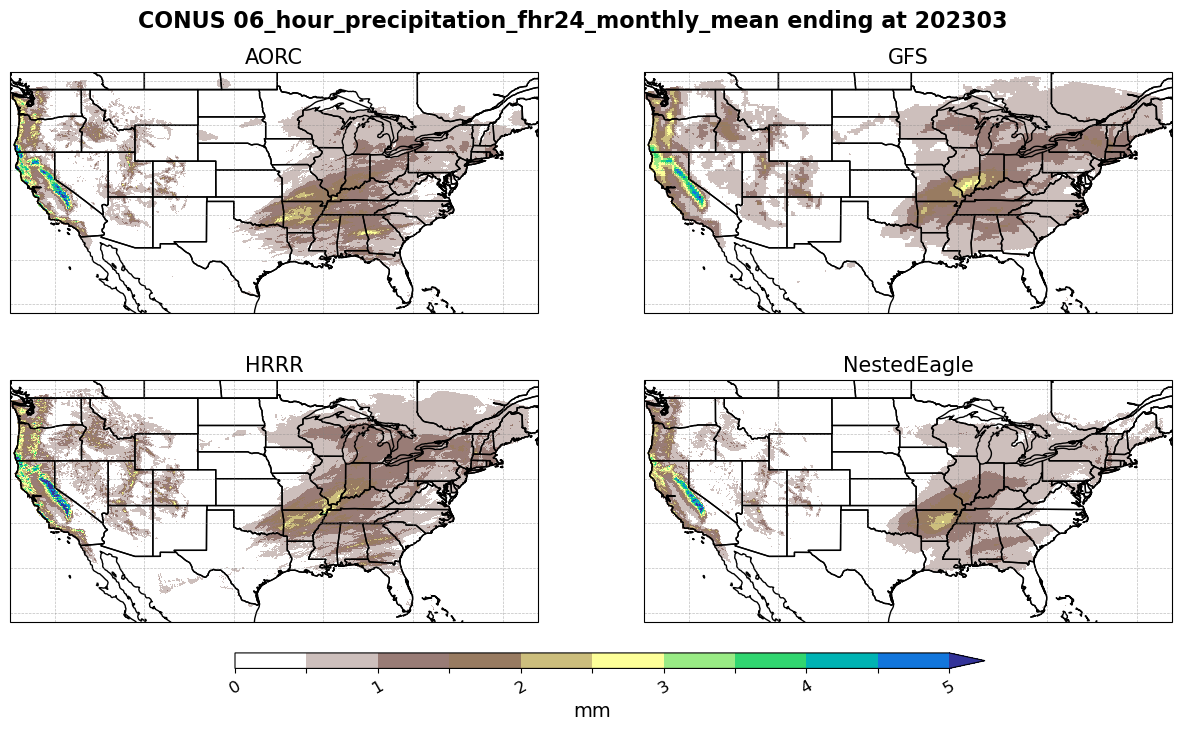}
    \caption{
        \textbf{Monthly mean 6~hour accumulated precipitation, March 2023.}
        Produced with a 24~hour lead time and valid during March 2023.
    }
    \label{sifig:precip-2023-03}
\end{figure}

\begin{figure}[H]
    \centering
    \includegraphics[
        width=\textwidth,
        trim=0 0 0 32,
        clip,
    ]{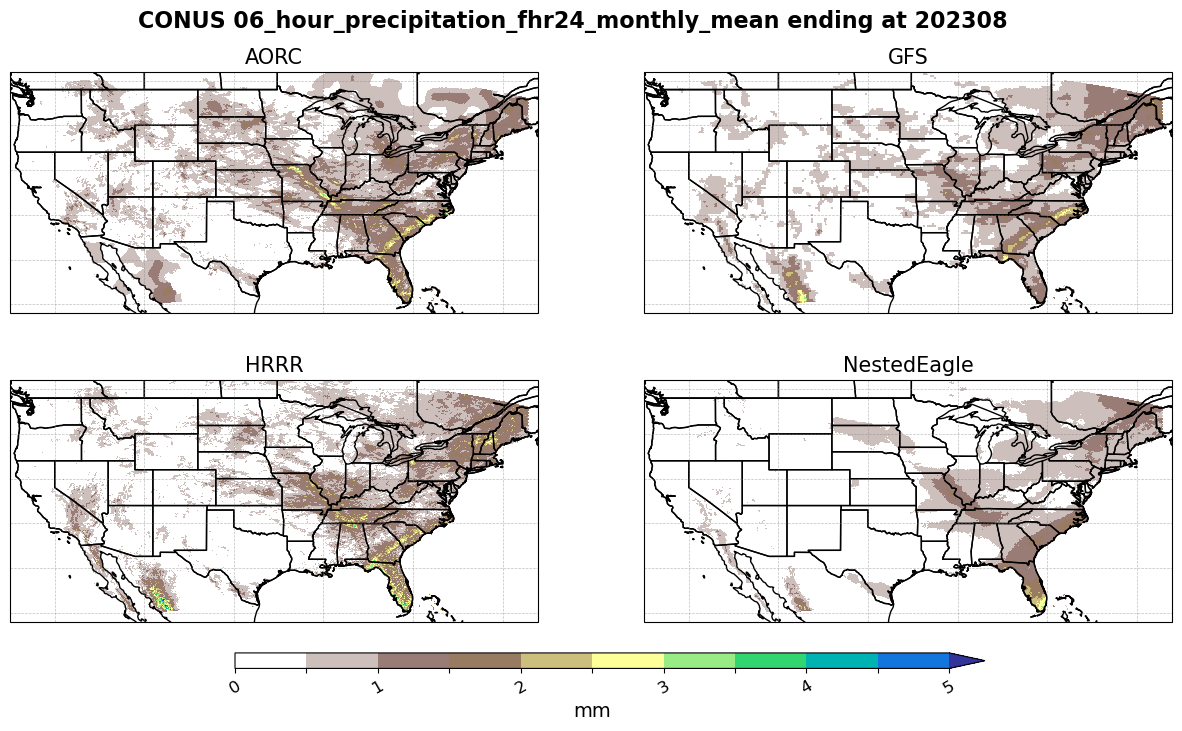}
    \caption{
        \textbf{Monthly mean 6~hour accumulated precipitation, August 2023.}
        Produced with a 24~hour lead time and valid during August 2023.
    }
    \label{sifig:precip-2023-08}
\end{figure}

Here we show monthly mean precipitation forecast visualizations that qualitatively support the
findings in \cref{subsec:fss_percentile}.
Recall that \cref{subsec:fss_percentile} shows the Fractions Skill Score
as a function of the percentile of each model's resolved amplitudes, and the
results indicate that the ML models tend to place
precipitation features in the right location, even if the amplitudes are
diminished.
\cref{sifig:precip-2023-03,sifig:precip-2023-08}
show monthly mean precipitation forecasts valid during March and August
2023 (i.e., from the validation set).
The quantity shown in each figure
is a monthly mean 6~hour precipitation accumulation, produced at a 24~hour
lead time.

Throughout the validation period, Nested-EAGLE and HRRR show qualitatively
similar skill in terms of their ability to place precipitation events over
CONUS, apart from the muted extrema in Nested-EAGLE.
However, during the months shown, Nested-EAGLE places several features noticeably
better, using the Analysis of Record for Calibration (AORC) dataset as a reference.
During March 2023, Nested-EAGLE properly places strong precipitation over
Arkansas,
whereas HRRR places the stronger precipitation farther northeast, on
the border of Missouri, Kentucky, and Tennessee.
During August 2023 we see similar behavior, where Nested-EAGLE and AORC show
strong precipitation extending across Missouri and into Tennessee, whereas HRRR
appears to miss the full extent of this feature.

\clearpage
\section{Sample Forecasts}
\label{si:sample-forecasts}

Here we show sample forecasts for select quantities.
The forecasts highlight a particularly strong atmospheric river event that
struck the West Coast of the United States on approximately 10 March 2023.
Not only is this event useful because of its societal relevance, but it is also an
excellent case for qualitative evaluations of how features are resolved across
the GFS--HRRR boundary.

% The lead time labels and the colorbar units are annotated here rather than
% baked into the images, so they stay editable. Both label nodes are drawn
% [overlay] so they do not widen or deepen the picture: the lead times sit in
% the page whitespace left of the 0.7\textwidth panels, and the units sit in
% the strip below the colorbar of the last panel. \fcstpanel takes the
% \includegraphics trim, the image file, the lead time in hours, and the valid
% time. The label node needs an align key: a node without one typesets its text
% in LR mode, where the \\ between the two lines is illegal.
\newcommand{\fcstpanel}[4]{%
    \begin{tikzpicture}
        \node[anchor=south west,inner sep=0] (img)
            {\includegraphics[width=.7\textwidth,trim=#1,clip]{#2}};
        \node[overlay,anchor=east,align=center,inner sep=0,font=\scriptsize]
            at ([xshift=-2pt]img.west) {$t_0 + #3$~hours\\#4};
    \end{tikzpicture}%
}

\begin{figure}[H]
    \centering
    \fcstpanel{0 105 0 5}{supplemental/forecasts/10m_wind_speed.2023-03-08T06.2023-03-08T12.jpeg}{6}{1200 UTC 8 March}
    \fcstpanel{0 105 0 25}{supplemental/forecasts/10m_wind_speed.2023-03-08T06.2023-03-09T06.jpeg}{24}{0600 UTC 9 March}
    \fcstpanel{0 105 0 25}{supplemental/forecasts/10m_wind_speed.2023-03-08T06.2023-03-10T06.jpeg}{48}{0600 UTC 10 March}
    \begin{tikzpicture}
        \node[anchor=south west,inner sep=0] (img)
            {\includegraphics[width=.7\textwidth,trim=0 55 0 25, clip]{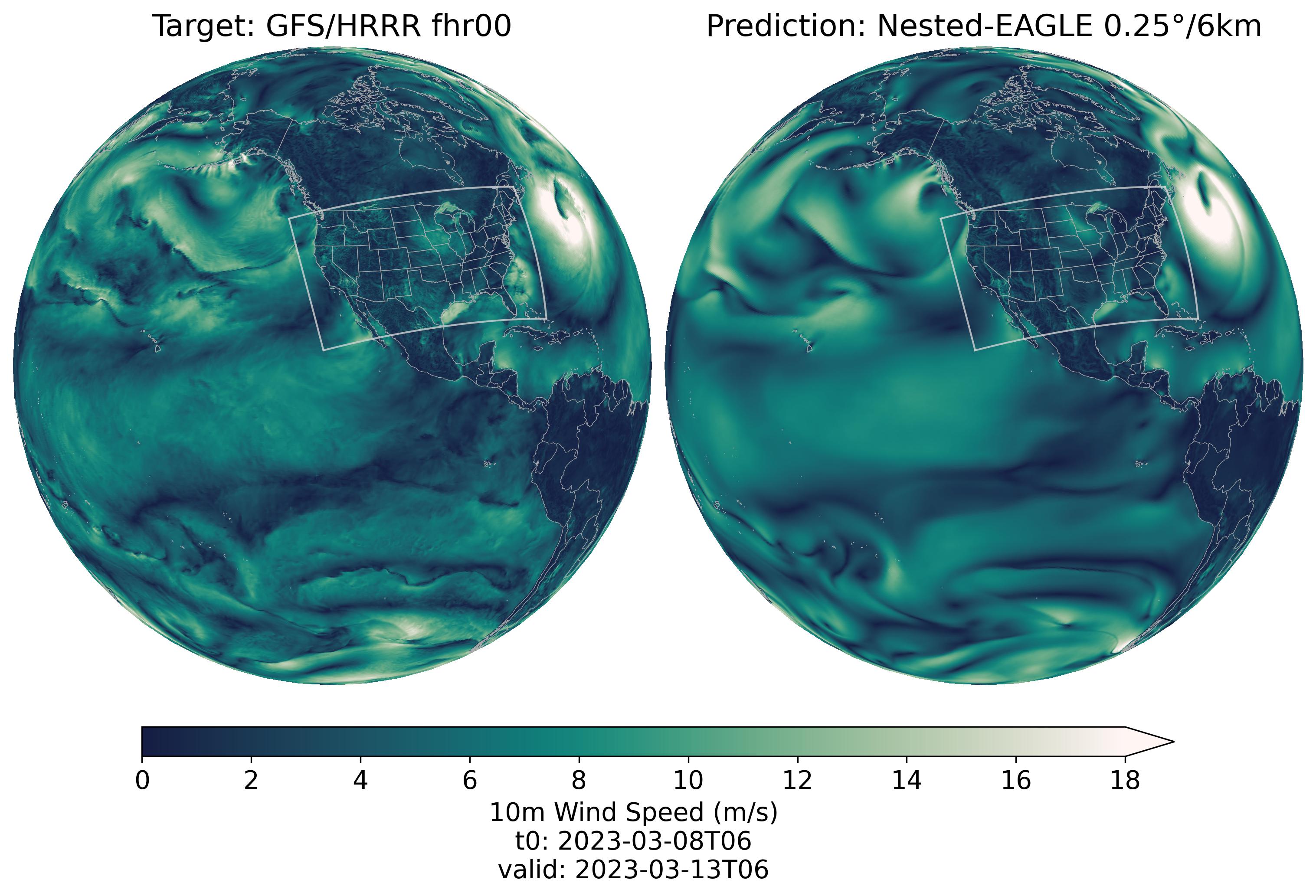}};
        \node[overlay,anchor=east,align=center,inner sep=0,font=\scriptsize]
            at ([xshift=-2pt]img.west) {$t_0 + 120$~hours\\0600 UTC 13 March};
        % Not [overlay]: the colorbar tick labels run to the very bottom of the
        % trimmed image, so this node has to grow the picture downward and push
        % the caption clear of it.
        \node[anchor=north,inner sep=0,font=\footnotesize]
            at ([yshift=-1pt]img.south) {m s$^{-1}$};
    \end{tikzpicture}
    \caption{
        \textbf{10m wind speed}. $t_0$ = 0600 UTC 8 March 2023.
    }
    \label{sifig:10m-wind-speed}
\end{figure}

\begin{figure}[H]
    \centering
    \fcstpanel{0 105 0 5}{supplemental/forecasts/2m_temperature.2023-03-08T06.2023-03-08T12.jpeg}{6}{1200 UTC 8 March}
    \fcstpanel{0 105 0 25}{supplemental/forecasts/2m_temperature.2023-03-08T06.2023-03-09T06.jpeg}{24}{0600 UTC 9 March}
    \fcstpanel{0 105 0 25}{supplemental/forecasts/2m_temperature.2023-03-08T06.2023-03-10T06.jpeg}{48}{0600 UTC 10 March}
    \begin{tikzpicture}
        \node[anchor=south west,inner sep=0] (img)
            {\includegraphics[width=.7\textwidth,trim=0 55 0 25, clip]{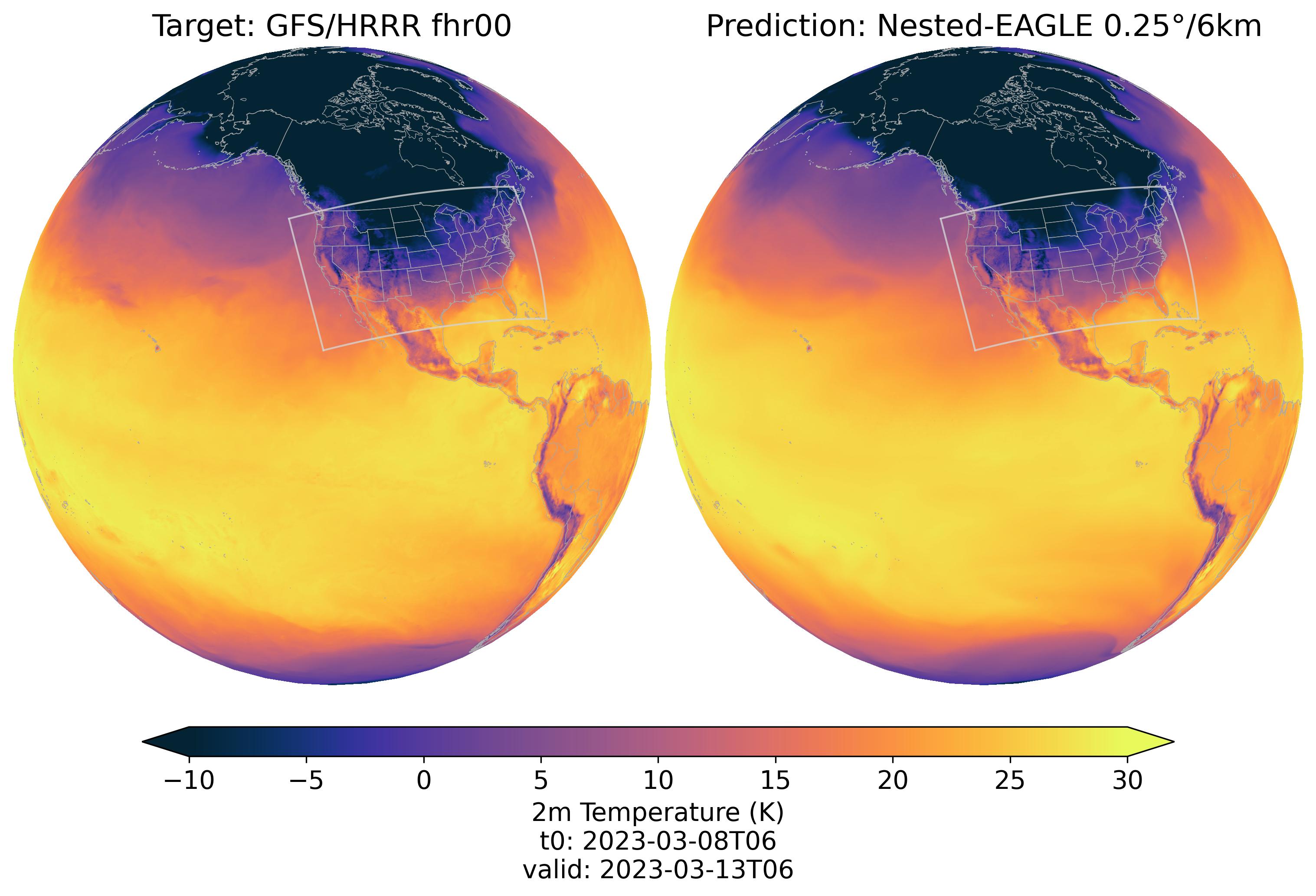}};
        \node[overlay,anchor=east,align=center,inner sep=0,font=\scriptsize]
            at ([xshift=-2pt]img.west) {$t_0 + 120$~hours\\0600 UTC 13 March};
        % Not [overlay]: the colorbar tick labels run to the very bottom of the
        % trimmed image, so this node has to grow the picture downward and push
        % the caption clear of it.
        \node[anchor=north,inner sep=0,font=\footnotesize]
            at ([yshift=-1pt]img.south) {$^\circ$C};
    \end{tikzpicture}
    \caption{
        \textbf{2m temperature}. $t_0$ = 0600 UTC 8 March 2023.
    }
    \label{sifig:2m-temperature}
\end{figure}

\begin{figure}[H]
    \centering
    \fcstpanel{0 105 0 5}{supplemental/forecasts/2m_specific_humidity.2023-03-08T06.2023-03-08T12.jpeg}{6}{1200 UTC 8 March}
    \fcstpanel{0 105 0 25}{supplemental/forecasts/2m_specific_humidity.2023-03-08T06.2023-03-09T06.jpeg}{24}{0600 UTC 9 March}
    \fcstpanel{0 105 0 25}{supplemental/forecasts/2m_specific_humidity.2023-03-08T06.2023-03-10T06.jpeg}{48}{0600 UTC 10 March}
    \begin{tikzpicture}
        \node[anchor=south west,inner sep=0] (img)
            {\includegraphics[width=.7\textwidth,trim=0 55 0 25, clip]{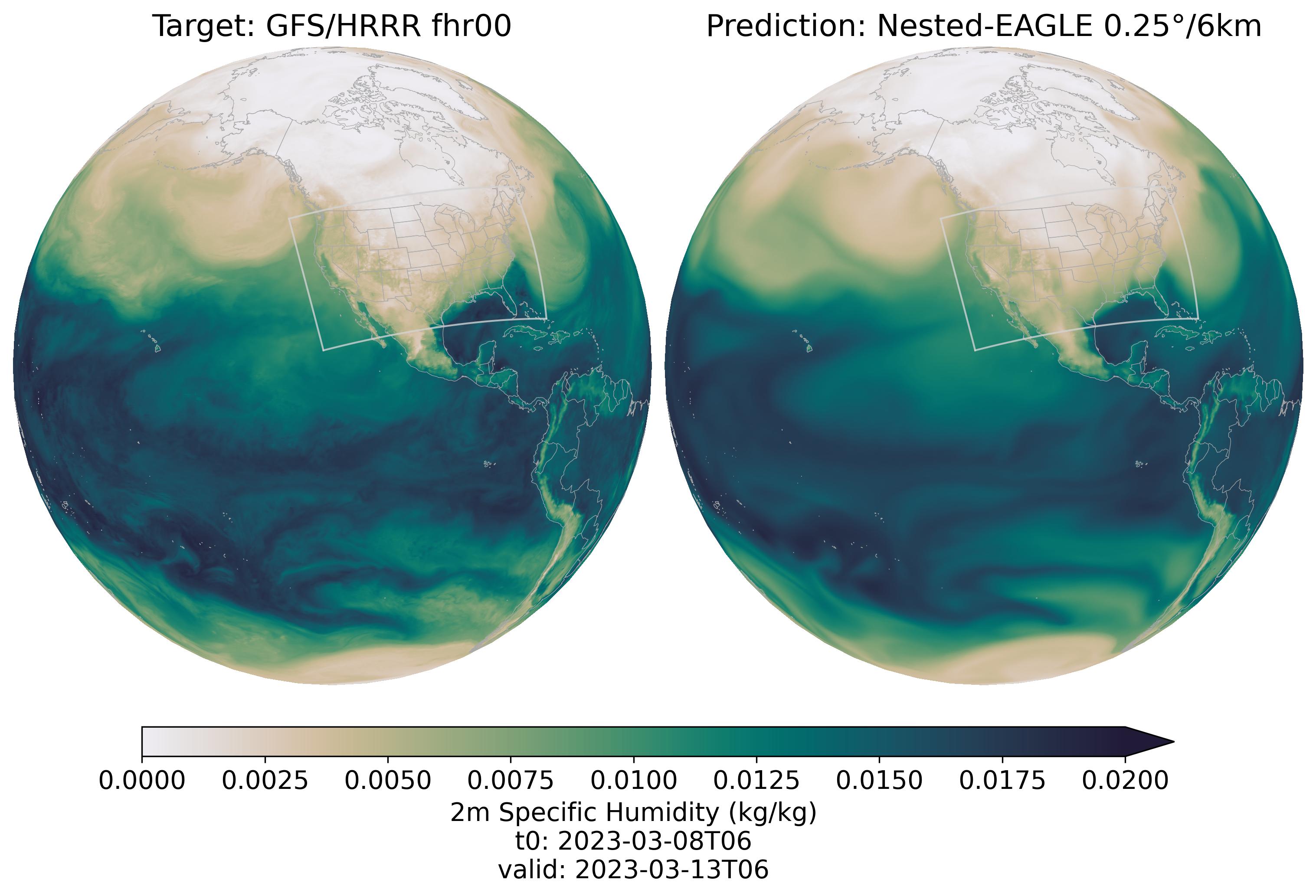}};
        \node[overlay,anchor=east,align=center,inner sep=0,font=\scriptsize]
            at ([xshift=-2pt]img.west) {$t_0 + 120$~hours\\0600 UTC 13 March};
        % Not [overlay]: the colorbar tick labels run to the very bottom of the
        % trimmed image, so this node has to grow the picture downward and push
        % the caption clear of it.
        \node[anchor=north,inner sep=0,font=\footnotesize]
            at ([yshift=-1pt]img.south) {kg kg$^{-1}$};
    \end{tikzpicture}
    \caption{
        \textbf{2m specific humidity}. $t_0$ = 0600 UTC 8 March 2023.
    }
    \label{sifig:2m-specific-humidity}
\end{figure}

\begin{figure}[H]
    \centering
    \fcstpanel{0 105 0 5}{supplemental/forecasts/total_precipitation_6hr.2023-03-08T06.2023-03-08T12.jpeg}{6}{1200 UTC 8 March}
    \fcstpanel{0 105 0 25}{supplemental/forecasts/total_precipitation_6hr.2023-03-08T06.2023-03-09T06.jpeg}{24}{0600 UTC 9 March}
    \fcstpanel{0 105 0 25}{supplemental/forecasts/total_precipitation_6hr.2023-03-08T06.2023-03-10T06.jpeg}{48}{0600 UTC 10 March}
    \begin{tikzpicture}
        \node[anchor=south west,inner sep=0] (img)
            {\includegraphics[width=.7\textwidth,trim=0 55 0 25, clip]{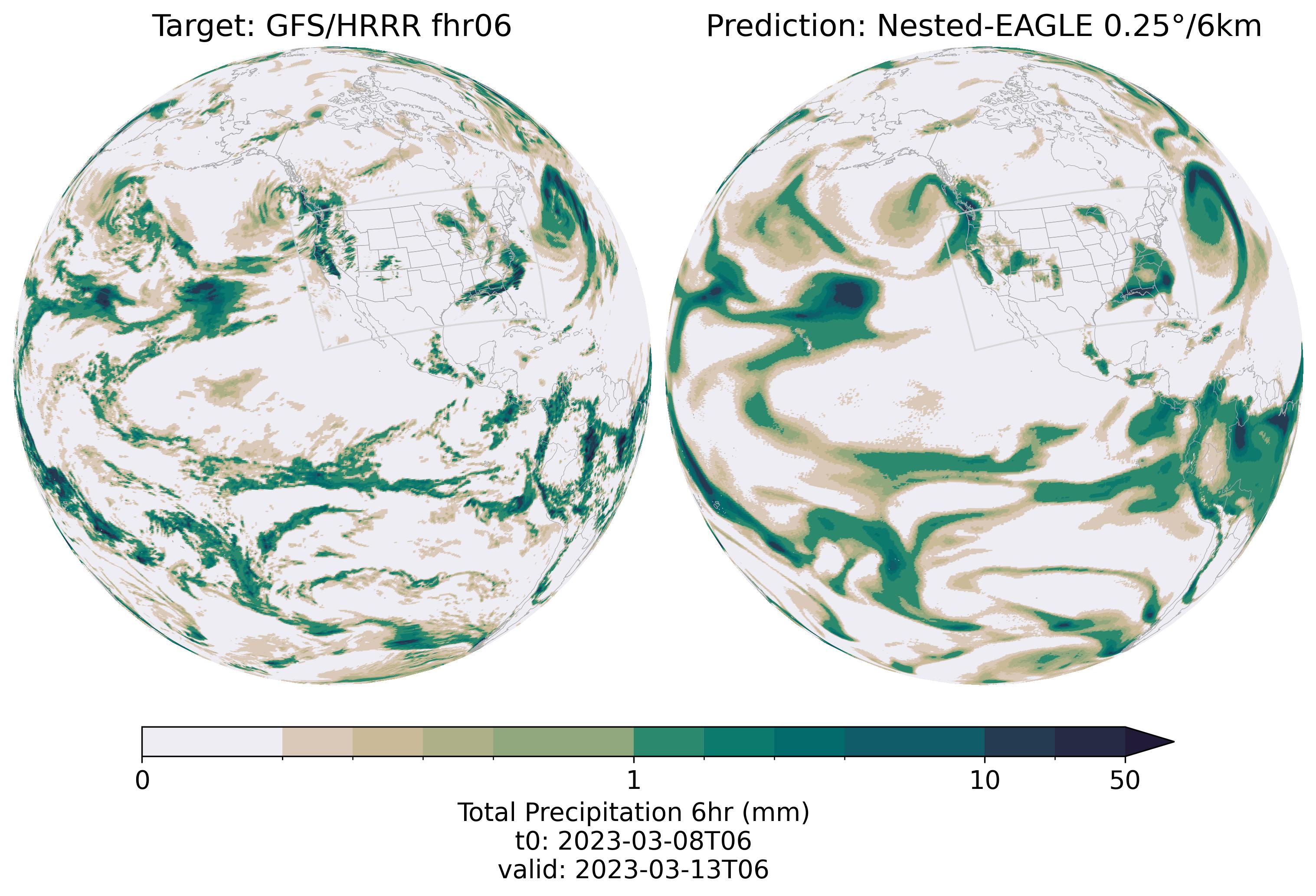}};
        \node[overlay,anchor=east,align=center,inner sep=0,font=\scriptsize]
            at ([xshift=-2pt]img.west) {$t_0 + 120$~hours\\0600 UTC 13 March};
        % Not [overlay]: the colorbar tick labels run to the very bottom of the
        % trimmed image, so this node has to grow the picture downward and push
        % the caption clear of it.
        \node[anchor=north,inner sep=0,font=\footnotesize]
            at ([yshift=-1pt]img.south) {mm / 6h};
    \end{tikzpicture}
    \caption{
        \textbf{Accumulated precipitation}. $t_0$ = 0600 UTC 8 March 2023.
    }
    \label{sifig:precip-forecast}
\end{figure}

\stopcontents[si]

\end{document}